\documentclass[onecolumn,11pt,table]{IEEEtran}

\usepackage[cmex10,intlimits]{amsmath}

\usepackage{amsfonts}
\usepackage{amssymb} 

\usepackage{calc}
\usepackage{circuitikz}
\usepackage{tikz}
\usepackage{wrapfig}
\usepackage{pgfplots}
\usepackage{booktabs}
\usepackage{multirow}
\usetikzlibrary{mindmap,trees}
\usepackage[normalem]{ulem}
\usepackage[nolist]{acronym}
\usepackage{subcaption}
\usetikzlibrary{%
  shapes.misc,% wg. rounded rectangle
  shapes.arrows,%
  chains,%
  matrix,%
  positioning,% wg. " of "
  scopes,%
  decorations.pathmorphing,% /pgf/decoration/random steps | erste Graphik
  shadows%
}

\usepackage[colorlinks]{hyperref}
\usepackage{xcolor}
\hypersetup{
colorlinks,
    linkcolor={red!20!black},
    citecolor={blue!20!black},
    urlcolor={blue!20!black}
}

\usepackage[most]{tcolorbox}

\usepackage{xparse}

\newtcolorbox[auto counter]{boxexampl}[2][]{floatplacement=t,float,%
colback=blue!0!white,colframe=blue!75!black,fonttitle=\bfseries,fontupper=\small,title=Box~\thetcbcounter. #2,#1}

\usepackage[colorinlistoftodos,prependcaption,textsize=small]{todonotes}
\usepackage{regexpatch}
\makeatletter
\xpatchcmd{\@todo}{\setkeys{todonotes}{#1}}{\setkeys{todonotes}{inline,#1}}{}{}
\makeatother

\colorlet{F1}{blue!8}
\colorlet{F2}{blue!4}
\colorlet{C1}{green!8}
\colorlet{C2}{green!4}
\colorlet{S1}{red!8}
\colorlet{S2}{red!4}
\colorlet{Q1}{gray!8}
\colorlet{Q2}{gray!4}

\providecommand*{\eu}{\ensuremath{\mrm{e}}}

\providecommand*{\ju}{\ensuremath{\mrm{j}}}
\newcommand{\ie}{\textit{i.e.}\/, }
\newcommand{\eg}{\textit{e.g.}\/, }
\newcommand{\cf}{\textit{cf.}\/ }
\providecommand*{\unit}[1]{\ensuremath{\mrm{\,#1}}}  % for units, eg $\unit{m}$
\renewcommand{\vec}[1]{{\boldsymbol#1}}
\providecommand*{\mat}[1]{\mathbf#1}
\providecommand*{\mrm}[1]{\mathrm{#1}}
\renewcommand{\Re}{\operatorname{Re}}	% The LaTeX standard is not ISO! 
\renewcommand{\Im}{\operatorname{Im}}	% The LaTeX standard is not ISO!
\newcommand{\partder}[2]{\frac{\partial #1}{\partial #2}}
\newcommand{\R}{\mathbb{R}}
\newcommand{\Pv}{\vec{P}}
\newcommand{\Ev}{\vec{E}}
\newcommand{\Hv}{\vec{H}}
\newcommand{\rvh}{\hat{\vec{r}}}
\newcommand{\Fv}{\vec{F}}

\newcommand{\Jv}{\vec{J}}
\newcommand{\rv}{\vec{r}}
\newcommand{\Xm}{\mat{X}}
\newcommand{\Rm}{\mat{R}}
\newcommand{\Jm}{\mat{I}}

\newcommand{\Bm}{\mat{B}}
\newcommand{\Lm}{\mat{L}}
\newcommand{\Vm}{\mat{V}}
\newcommand{\Zm}{\mat{Z}}
\newcommand{\Cm}{\mat{C}}
\newcommand{\Om}{\mat{0}}
\newcommand{\Cmi}{\mat{C}_{\mrm{i}}}
\newcommand{\Um}{\mat{U}}
\newcommand{\Zin}{Z_{\mrm{in}}}
\newcommand{\Xin}{X_{\mrm{in}}}
\newcommand{\Rin}{R_{\mrm{in}}}
\newcommand{\Vin}{V_{\mrm{in}}}
\newcommand{\Iin}{I_{\mrm{in}}}
\newcommand{\refl}{\varGamma}  % reflection coefficient
\newcommand{\band}{B}  % fractional bandwidth
\newcommand{\QFBW}{Q_{\mrm{FBW}}}  % Q-fractional bandwidth
\newcommand{\QZp}{Q_{\mrm{Z}'}}  % Q- from Zin'
\newcommand{\Id}{\mat{1}}					
\providecommand*{\diff}{\operatorname{d}\!}

\providecommand*{\diffV}{\operatorname{dV}\!}
\newcommand{\trans}{\text{T}}
\newcommand{\herm}{\text{H}}
\def\checkmark{\tikz\fill[scale=0.4](0,.35) -- (.25,0) -- (1,.7) -- (.25,.15) -- cycle;} 
\newcommand{\ok}{\checkmark}
\newcommand{\no}{}%{no}

\newcommand{\omegar}{\omega _{\mathrm{r}}}
\newcommand{\omegap}{\omega _{\mathrm{p}}}
\newcommand{\tdv}[1]{\boldsymbol{\mathcal{#1}}}	% define time domain style here
\newcommand{\Etd}{\tdv{E}}
\newcommand{\Htd}{\tdv{H}}
\newcommand{\Jtd}{\tdv{J}}
\newcommand{\Ptd}{\tdv{P}}
\newcommand{\Wtd}{\mathcal{W}}

\newcommand{\WDISS}{W_\mathrm{diss}}
\newcommand{\WEM}{W_\mathrm{EM}}
\newcommand{\WSTO}{W_\mathrm{sto}}
\newcommand{\WREC}{W_\mathrm{rec}}
\newcommand{\Wreac}{W_{\mathrm{reac}}}
\newcommand{\Wem}{\Wtd_\mathrm{EM}}
\newcommand{\Wsto}{\Wtd_\mathrm{sto}}
\newcommand{\Wrec}{\Wtd_\mathrm{rec}}
\newcommand{\WXp}{W_{\Xm'}}

\providecommand*{\V}[1]{\boldsymbol{#1}} % vector (bold italics)
\providecommand*{\D}[1]{\,\mathrm{d}#1}  % differential (small space + upright)

\providecommand*{\EPS}{\varepsilon} % permittivity
\providecommand*{\MUE}{\mu} 		% permeability
\providecommand*{\SIGMA}{\sigma} 	% conductivity

\providecommand*{\Wsto}{W_\mathrm{sto}}
\ctikzset{bipoles/length=0.9cm}

\newacro{MoM}[MoM]{method of moments}
\newacro{PEC}[PEC]{perfect electric conductor}
\newacro{EFIE}[EFIE]{electric field integral equation}
\newacro{MFIE}[MFIE]{magnetic field integral equation}
\newacro{FBW}[FBW]{fractional bandwidth}
\newacro{RFID}[RFID]{Radio Frequency Identification}
\newacro{LTATDMS}[LTATDMS]{long test acronym that doesn't make sense}

\usepackage{etoolbox}
\newtoggle{fullpaper}
\toggletrue{fullpaper}

\begin{document}
\title{Energy Stored by Radiating Systems}
\author{Kurt~Schab,  Lukas~Jelinek, Miloslav~Capek, Casimir~Ehrenborg, \\Doruk~Tayli, Guy~A.~E.~Vandenbosch, and Mats~Gustafsson
\thanks{K.~Schab is with the Department of Electrical and Computer
Engineering, Antennas and Electromagnetics Laboratory, North Carolina State University, Raleigh, NC, USA (e-mail: krschab@ncsu.edu).}
	\thanks{L.~Jelinek and M.~Capek are with the Department of Electromagnetic Field, Faculty of Electrical Engineering, Czech Technical University in Prague, Technicka~2, 16627, Prague, Czech Republic
(e-mail: \mbox{lukas.jelinek@fel.cvut.cz}, \mbox{miloslav.capek@fel.cvut.cz}).}%
\thanks{C. Ehrenborg, D. Tayli, and M. Gustafsson are with the Department of Electrical and Information Technology, Lund University, Box 118, SE-221 00 Lund, Sweden. (Email: {casimir.ehrenborg,doruk.tayli,mats.gustafsson\}@eit.lth.se).}}%
\thanks{G.~A.~E.~Vandenbosch is with the Department of Electrical Engineering, KU Leuven, Leuven, Belgium. (Email: guy.vandenbosch@kuleuven.be).}%
\thanks{This work was supported by the Czech Science Foundation under project \mbox{No.~GJ15-10280Y}, Swedish Foundation for Strategic Research for the project Complex analysis and convex optimization for EM design, and the United States Intelligence Community Postdoctoral Research Fellowship Program.}
}
%\markboth{Journal of \LaTeX\ Class Files,~Vol.~XX, No.~XX, \today}%
%{Mickey \MakeLowercase{\textit{et al.}}: TBD}

\maketitle

\begin{abstract}
	Though commonly used to calculate Q-factor and fractional bandwidth, the energy stored by radiating systems (antennas) is a subtle and challenging concept that has perplexed researchers for over half a century. Here, the obstacles in defining and calculating stored energy in general electromagnetic systems are presented from first principles as well as using demonstrative examples from electrostatics, circuits, and radiating systems.  Along the way, the concept of unobservable energy is introduced to formalize such challenges.  Existing methods of defining stored energy in radiating systems are then reviewed in a framework based on technical commonalities rather than chronological order. Equivalences between some methods under common assumptions are highlighted, along with the strengths, weaknesses, and unique applications of certain techniques.  Numerical examples are provided to compare the relative margin between methods on several radiating structures.
\end{abstract}


\IEEEpeerreviewmaketitle

\section{Introduction}
For many in the field of electromagnetics, stored energy is best known by its appearance in the definition of a time-harmonic system's Q-factor (quality factor, antenna Q, radiation Q) \cite{IEEEStd_antennas, IEEEStd_filters},
\begin{equation}
\label{eq:qdef}
Q = \frac{2\pi \WSTO}{\WDISS},
\end{equation}
from which an estimate of fractional bandwidth is available.  In the above expression, $\WSTO$ and $\WDISS$ denote the cycle-mean stored and dissipated energies within the system, respectively.  The dissipated energy is typically well defined and can be easily calculated, while in many cases the definition of stored energy is ambiguous.  This issue is particularly troublesome in distributed and radiating systems, where there exists no consistent, physically-intuitive method of delineating the overlap between energy which is stored and that which is propagating.  Analogous problems can be encountered in lumped circuits, where specific networks can be arbitrarily inserted to increase the total energy without altering the impedance characteristics as seen from a port. The first of two goals of this paper is to elucidate the challenges involved in defining stored energy within a general electromagnetic system.  To do so, we draw upon examples of lumped circuits and radiating systems which exhibit the general issue of ``unobservable energy states''.  Although this concept is somewhat abstract, it provides a consistent framework for understanding what makes defining stored energy in certain systems so difficult.

Because of the powerful relationship between fractional bandwidth and stored energy, many researchers have worked to rigorously define stored energy in an attempt to obtain bounds on the broadband behavior of systems.  Of particular practical and historical importance is the study of stored energy in radiating systems, \ie antennas.  Work in this area dates back over half a century and has given rise to many unique (and sometimes controversial) interpretations and claims.  One regime where most methods agree is in the quasi-static limit, \ie for small antennas.   However, for problems involving larger antennas or antennas next to larger objects (\eg ground planes or human bodies), most methods disagree and there is no consensus on a definition of stored energy. In some cases, the similarities and differences between these existing approaches are clear, though in other instances the technical and philosophical connections between works from different eras are more subtle.  The second goal of this paper is to provide a clear summary of the many previously published approaches to defining stored energy, with emphasis on works studying distributed and radiating systems.  We aim to provide not a chronological history of this topic, but rather an organized guide to the major themes and concepts used in previous works. 

The paper is organized as follows.  In Section~\ref{sec:SEdef}, we present a general definition for stored energy within an electromagnetic system using the concept of unobservable energy states.  In Section~\ref{SecSto}, existing approaches to defining and calculating stored energy within radiating systems are summarized.  Where applicable, the similarities and differences between these methods are highlighted, along with their strengths, weaknesses, and relation to the formal definition of stored energy given in Section~\ref{sec:SEdef}.  Analytical and numerical examples are presented in Section~\ref{sec:ANCompar}, giving both quantitative and qualitative insight into the relative results obtained by the methods outlined in Section~\ref{SecSto}.  The paper concludes with a discussion of applications of certain methods in Section~\ref{sec:Applications} and general conclusions in Section~\ref{sec:discussion}. Further details are provided on the classical definition of stored energy in Box~\ref{B:storedEnergy}, unobservable states in Boxes~\ref{B:UnObsWa} and~\ref{B:UnObsWb}, and electrostatic energy in Box~\ref{B:Electrostatics}.

\section{Definition and physical rationale of stored EM energy}
\label{sec:SEdef}
\iftoggle{fullpaper}{%

\begin{boxexampl}[label={B:storedEnergy}]{Stored energy in circuits and systems}
Many dynamic systems in nature can be modeled as 
\begin{equation}
  \partder{}{t}\mat{W}\mat{u}+\mat{P}\mat{u}=\mat{B}\mat{v}_{\mrm{in}}\quad\text{with}\quad
	\mat{u}_{\mrm{out}} = \mat{B}^{\trans}\mat{u},
\label{eq:Sysmodel}
\end{equation}
where $\mat{v}_{\mrm{in}}$ and $\mat{u}_{\mrm{out}}$ denote the input and output signals, $\mat{u}$ the system's internal states, and $\mat{W}$, $\mat{P}$, and $\mat{B}$ are matrices describing the system~\cite{Willems1972b}. To construct an energy balance of such a system over an interval $[t_1,t_2]$ we multiply with the states $\mat{u}$ and integrate to get
\begin{equation}
  \left[\frac{\mat{u}^{\trans}\mat{W}\mat{u}}{2}\right]_{t_1}^{t_2}
  +\int_{t_1}^{t_2}\mat{u}^{\trans}\mat{P}\mat{u}\diff t
  =\int_{t_1}^{t_2} \mat{u}_{\mrm{out}}^{\trans}\mat{v}_{\mrm{in}}\diff t,
\label{eq:SysEnDef}
\end{equation} 
in which $^\trans$ denotes matrix transpose. The left-hand side can be identified as the difference in stored energy and dissipation of energy during the interval and the right-hand side is the supplied energy, \cf the definition in Section~\ref{sec:SEdef}. The definition and interpretation of the stored energy depend on the properties of the matrices $\mat{W}$, $\mat{P}$, and $\mat{B}$.

Systems representable by~\eqref{eq:Sysmodel} can contain states that are unobservable to an observer seeing only the input and output signals. These states can contain unobservable energy~\cite{Willems1972b}.  The time-average stored energy~\eqref{eq:SysEnDef} for time-harmonic signals $\mat{u}(t)=\Re\{\mat{U}\eu^{\ju\omega t}\}$ is $\mat{U}^{\herm}\mat{W}\mat{U}/4$, where we note that the system matrix $\mat{W}$ can be determined by frequency differentiation of the matrix $\Zm$ obtained from \eqref{eq:Sysmodel}, \ie
\begin{equation}
  \Zm = \mat{P} + \ju\omega\mat{W} \quad \text{with} \quad \mat{W} = \partder{\Im\{\Zm\}}{\omega}.
\label{eq:Zstatespace}
\end{equation}
By \eqref{eq:Sysmodel}, it is implicit that $\mat{P}$ and $\mat{W}$ are frequency-independent in this classical system model. Probably one of the most familiar systems which follows the form~\eqref{eq:SysEnDef} is a lumped circuit. Here, the input and output states are the voltages $\mat{V}$ and currents $\mat{I}$, respectively. These are related through either the explicit summation of all circuit components or their impedance matrix~\cite{Smith1969} 
\begin{equation}
	\Zm = \Rm + \ju\omega\Lm + \frac{1}{\ju\omega}\Cmi,
\label{eq:ImpMat}
\end{equation}
where $\Rm$ describes the resistive components of the circuit and matrices $\Lm$ and $\Cmi$ represent the reactive elements. The impedance matrix relates the current to the voltage as $\mat{Z}\mat{I}=\mat{V}$. To reach the stored energy form in~\eqref{eq:SysEnDef} we differentiate the impedance matrix with respect to $\omega$ and multiply with the current $\Jm$ and its hermitian conjugate $\Jm^{\herm}$ from the right and left, respectively. This expresses the time-average stored energy, average of the first term in~\eqref{eq:SysEnDef} for a time-harmonic signal, as the quadratic form~\cite{Smith1969}
\begin{equation}
	\WSTO = \frac{1}{4}\Jm^{\herm}\Lm\Jm + \frac{1}{4\omega^2}\Jm^{\herm}\Cmi\Jm,
\label{eq:}
\end{equation}  
where the classical expressions for the stored energy in inductors and capacitors are recognized~\cite{Wing2008}. 
\end{boxexampl}
}{}

The total energy of a dynamic system, see Box~\ref{B:storedEnergy}, represents a well-known and fundamental characteristic describing the energy stored in all of its degrees of freedom. By contrast, the observable part of total energy is a more subtle quantity typically defined in such a way that its value has a direct correspondence with the input / output relation of the system as seen by a fixed observer~\cite{Willems1972b}. In lossless systems, these two quantities are equal due to the Foster's reactance theorem~\cite[Sec.~8-4]{Harrington_TimeHarmonicElmagField}. In general dissipative systems, however, they lose their relation due to the presence of states not observable from outside the system, see Boxes~\ref{B:UnObsWa} and~\ref{B:UnObsWb}.

\begin{boxexampl}[label={B:UnObsWa}]{{Unobservable energy, part 1}}
{\centering
  \includegraphics[width=0.65\textwidth]{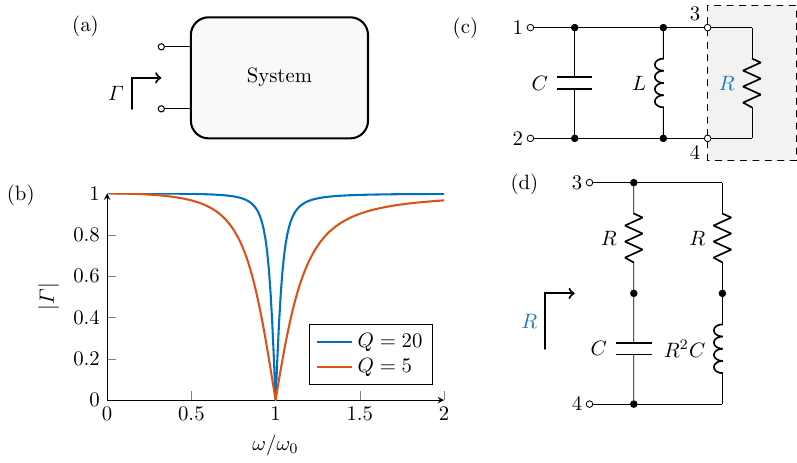}\par}
  
\textit{The unobservable states are defined as those states which cannot be identified by the observer.}
To provide an example, let us suppose a yet unknown system, schematically depicted in panel (a). This system is examined by an observer at its input port and quantified by its reflection coefficient $\varGamma$, \cite{Collin_FoundationsForMicrowaveEngineering}. From the information obtained at the port, we can attempt to construct the system within. The simplest circuit that fits the measured data, depicted in panel (b), is an RLC circuit, see panel (c). However, the resistor in the RLC circuit can be arbitrarily replaced by circuit elements of the Z\"obel type~\cite{1923_Zobel_BSTJ}, see panel (d), without affecting exterior results observed at the port. If we now assume to be able to access the internal structure of the constructed circuits, we can calculate the energy stored in the reactive elements. It then becomes apparent that the added Z\"obel circuit does affect the stored energy without changing what is observed at the port. Thus, these two valid circuit realizations for the same measured reflection coefficient predict different values of stored energy.
This illustrates that depending on the specific circuit realization, the stored energy, unlike the reflection coefficient, can potentially be altered by states \textit{unobservable} to the outside observer. This is true for all quantities inferred from stored energy, including the Q-factor in \eqref{eq:qdef}. It is also important to appreciate that how much of a system's stored energy is observable explicitly depends on the observer. If, for example, the observation procedure would include both measurement of the the reflection coefficient $\varGamma$ and measurement of heat produced by the circuit, the observer will be able to distinguish circuit (c) from circuit (d), since the time evolution of heat differs in them. %Such an observer will be able to observe to entirety of energy storage inside the circuit. Equivalently, the observer could enter the circuit and measure the voltages and currents over every circuit element.
%[\TR {The situation becomes much clearer when studying time domain. Consider a pure Z\"obel network excited by a very broad but finite rectangular current pulse. At any time within the time frame of the pulse, the power drawn from the current source is {$RI^2$}, just as for a pure resistance $R$. However, just after the pulse has started, this energy is (partially) used to build up the stored energy in the capacitors and inductors. After this starting up phase, a steady state regime is reached, where the inserted energy perfectly equals the energy dissipated in both resistors. This dissipation is identical as in the case of a pure resistor $R$. In the end phase, just after the pulse has stopped, and thus no further energy can be drawn from it, the energy stored in the capacitors and inductors is transferred to the resistors, where it is dissipated. As such, in time domain there is no violation whatsoever of the energy conservation law. The reactive elements only cause a shift in time between the power injected in the system and the dissipation in the resistors.}]
\end{boxexampl}

\begin{boxexampl}[label={B:UnObsWb}]{{Unobservable energy, part 2}}
{\centering
  \includegraphics[width=0.6\textwidth]{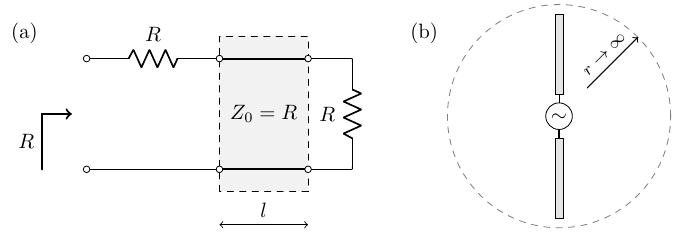}\par}

Unobservable energy can be encountered in many basic electromagnetic devices, such as a matched transmission line or a radiating antenna system, see panels (a) and (b) above. In both of these cases, traveling energy exists but is unobservable for an observer at the input port.  Specifically, the total energy within the transmission line circuit in panel (a) can be arbitrarily altered through changes to the line length $l$ with no effect on the impedance seen from the input port \cf with lumped circuit models for a transmission line~\cite{Feynman1965}. Similarly, the energy stored within the radiating system in panel (b) depends on the definition of the spatial boundary at which energy ``leaves'' the system, though this boundary has no effect on the port impedance.  For time-harmonic signals and a system boundary chosen at infinity, \ie the far-field sphere, the system in panel (b) contains an infinite amount of traveling energy.
\end{boxexampl}

\begin{figure}[]
\centering
\includegraphics[]{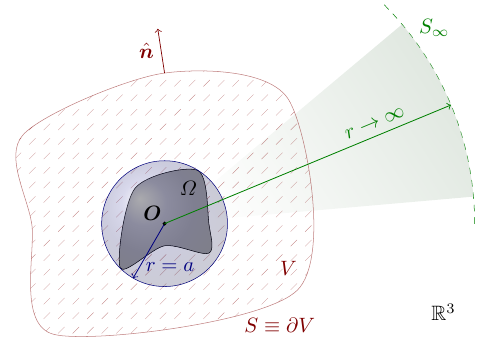}
\caption{Sketch of an antenna region $\varOmega$, a smallest circumscribing sphere of radius $a$, an arbitrary volume $V$ with its boundary surface $S$ and the far-field sphere bounded by $S_\infty$.}
\label{fig:FigBoundIntro}
\end{figure}

The energy supplied to a radiating system is converted into several different forms. Consider a radiator made of non-dispersive isotropic medium with permittivity $\EPS$, permeability $\MUE$ and conductivity $\SIGMA$, which is placed in otherwise free space (effects induced by frequency dispersion are discussed in Appendix~\ref{app:SEdefDisp}). The radiator is enclosed within a volume $V$ with bounding surface $S$, see Figure~\ref{fig:FigBoundIntro}.  Here we use, $\Etd$ and $\Htd$ to represent the time-domain electric and magnetic fields, respectively, while $\Jtd_\mathrm{source}$ denotes an impressed current distribution.  Assuming the initial conditions \mbox{$\Etd \left(\V{r}, t \to-\infty\right)=\V{0}$}, \mbox{$\Htd \left(\V{r}, t \to-\infty\right) = \V{0}$}, Poynting's theorem can be written as~\cite{Jackson1999,Ruppin_ElectromagneticEnergy}
\begin{equation}
  \label{eq:StEnDefEq1}
  \Wtd_\mathrm{supp} \left(t_0\right) = \Wtd_\mathrm{EM} \left(t_0\right) + \Wtd_\mathrm{heat} \left(t_0\right) + \Wtd_\mathrm{rad} \left(t_0\right),
\end{equation}
where the supplied energy is
\begin{equation}
  \label{eq:StEnDefEq2B}
  \Wtd_\mathrm{supp} \left(t_0\right) = -\int\limits_{-\infty}^{t_0} \int\limits_V \Etd\cdot\Jtd_\mathrm{source} \D{V} \D{t},
\end{equation}
the energy lost in heat is
\begin{equation}
  \label{eq:StEnDefEq2C}
  \Wtd_\mathrm{heat} \left(t_0\right) = \int\limits_{-\infty}^{t_0} \int\limits_V  \SIGMA \left| \Etd \right|^2  \D{V} \D{t},
\end{equation}
and the net energy escaping the volume through the bounding surface $S$ is
\begin{equation}
  \label{eq:StEnDefEq2D}
  \Wtd_\mathrm{rad} \left(t_0\right) = \int\limits_{-\infty}^{t_0} \int_S \left( \Etd \times \Htd \right) \cdot \hat{\boldsymbol{n}}\D{S} \D{t}.
\end{equation}
These terms account for energy supplied to and lost from the system, letting us define the remaining  term in Poynting's theorem as the total electromagnetic energy stored within the volume $V$ at time $t=t_0$,
\begin{equation}
  \label{eq:StEnDefEq2A}
  \Wtd_\mathrm{EM} \left(t_0\right) = \frac{1}{2} \int\limits_V \left(\EPS \left| \Etd \right|^2 + \MUE \left| \Htd \right|^2 \right) \D{V}.
\end{equation}
All aforementioned quantities depend upon a choice of volume $V$ and its bounding surface $S$. A specific choice of the surface $S$ lying in the radiation zone\footnote{Here we make an assumption that electric and magnetic fields are temporarily bandlimited and thus the radiation zone can be defined in a usual manner by the dominance of the $1/r$ field components.} \cite{Jackson1999} leads to \eqref{eq:StEnDefEq2A} representing the total electromagnetic energy and \eqref{eq:StEnDefEq2D} the total radiated energy. This division, however, depends on surface $S$ due to time retardation.

The energy defined in~\eqref{eq:StEnDefEq2A} encompasses all electromagnetic energy localized in the chosen volume $V$ containing the system. Nevertheless, for an observer situated at the input port of the system, the entirety of energy $W_\mathrm{EM}$ is not necessarily observable, see Box~\ref{B:UnObsWa}. Unobservable energy states by definition cannot affect physical measurements at the location of the observer. For this observer a more sensible definition of the stored energy is,
\begin{equation}
\label{eq:StEnDefEq3}
\Wtd_\mathrm{sto} \left(t_0\right) = \Wtd_\mathrm{EM} \left(t_0\right) - \Wtd_\mathrm{unobs} \left(t_0\right),
\end{equation}
where $\Wtd_\mathrm{unobs} \left(t_0\right)$ is the energy of all unobservable states. This definition suggests that the value of stored energy depends on the position of the observer. Throughout this paper it is assumed that the observer is positioned at the input port of the electromagnetic system and therefore perceives the minimum stored energy from all observers. Note, however, that the even the minimum value of energy $\Wtd_\mathrm{sto} \left(t_0\right)$ is not necessarily recoverable \cite{1970_Day_QJMAM,Polevoi_MaximumExtractableEnergy} by experiments performed at the location of the observer (recoverable energy $\Wrec \left(t_0\right)$ is detailed later in Section~\ref{SecSto}). The stored energy is fully recoverable only in special cases, the most important being closed lossless systems satisfying $\Wtd_\mathrm{heat} \left(t_0\right) + \Wtd_\mathrm{rad} \left(t_0\right) =0$. Examining the properties of aforementioned energy definitions, we arrive at the following inequality
\begin{equation}
\label{eq:StEnDefEq4}
0 \leq \Wrec \left(t_0\right) \leq \Wsto \left(t_0\right) \leq \Wem \left(t_0\right) \leq \Wtd_\mathrm{supp} \left(t_0\right).
\end{equation}

In the preceding discussion, all quantities are defined in the time domain.  However, in many cases cycle mean values of the energies in \eqref{eq:StEnDefEq2D}, \eqref{eq:StEnDefEq2A} and \eqref{eq:StEnDefEq3} in time-harmonic steady state are of interest, where time-harmonic quantities at angular frequency $\omega$ are defined as $\mathcal{G}(t) = \Re \{G(\omega)\eu^{\ju\omega t} \}$ and cycle means are denoted as $\langle\cdot\rangle$. The conversion of all preceding energy terms into time-harmonic domain is straightforward, but induces an issue with potentially unbounded energy values. This happens when the volume $V$ is chosen to consist of all space (denoted $V_\infty$) with bounding surface $S$ being a sphere at infinity (denoted $S_\infty$). In such a case the time-averaged total electromagnetic energy
\begin{equation}
\label{eq:StEnDefEq7}
\WEM = \left\langle {{\Wtd_{{\mathrm{EM}}}}} \right\rangle  
= \frac{1}{4}\int_{{V_\infty }} {\left( {\varepsilon {{\left| {\boldsymbol{E} \left( \omega \right)} \right|}^2} + \mu {{\left| {\boldsymbol{H} \left( \omega \right)} \right|}^2}} \right)\diff V}
\end{equation}
is infinite due to the infinite amount of radiation energy contained in propagating fields within the volume $V_\infty$. Subtracting this energy from the total energy $W_\mathrm{EM}$, \ie to identify unobservable energy with radiation, is the aim of several approaches calculating the stored energy $\WSTO = \left\langle \Wsto \left(t_0\right) \right\rangle$. These methods rely on the fact that time-averaged radiated power
\begin{equation}
\label{eq:StEnDefEq5}
P_\mathrm{rad}  
=\int_{S_\infty} \Pv(\omega) \cdot\rvh \D{S}
= \frac{1}{2Z_0}\int_{S_\infty } \left| {\boldsymbol{E} \left( \omega \right)} \right|^2\D{S}
= \frac{1}{2Z_0}\int_{S^2}  {\left| {\boldsymbol{F} \left( \omega \right)} \right|}^2\D{S}
\end{equation}
in time-harmonic steady state is the same for all surfaces enclosing the sources. The quantities
\begin{equation}
\label{eq:StEnDefEq6}
{\boldsymbol{F}}\left(\omega\right) = \mathop {\lim }\limits_{r \to \infty }  {r\eu^{\ju kr}\boldsymbol{E}(\omega)}
\quad\text{and}\quad
\Pv\left(\omega\right) = \frac{1}{2}\Re\{\Ev\left( \omega \right)\times\Hv^\ast\left( \omega \right)\}
\end{equation}
used above denote the far field and the real part of the Poynting's vector, respectively.  In the far right-hand-side of \eqref{eq:StEnDefEq5}, surface $S^2$ denotes the unit sphere and $k=\omega/c_0$ in~\eqref{eq:StEnDefEq6} denotes the free-space wavenumber.  When used to evaluate Q-factor, the cycle-mean stored energy $\WSTO$ is normalized by the cycle-mean dissipated energy (see \eqref{eq:qdef}).  In radiating systems without ohmic losses, the cycle-mean dissipation reduces to the radiated power $P_\mathrm{rad}$ in \eqref{eq:StEnDefEq5}.  

Note that in many cases, the Q-factor in \eqref{eq:qdef} is assumed to be tuned such that the system as a whole is resonant.  In general, a non-resonant system can be tuned by the addition of a specific reactance, which stores additional energy $W_\mathrm{tune}$.  The tuned Q-factor can then be explicitly rewritten as
\begin{equation}
\label{eq:tunedqdef}
Q = \frac{2\pi \left(\WSTO + W_\mathrm{tune}\right)}{\WDISS}.
\end{equation}
Since the stored energy in a pure reactance is well-defined, throughout this paper we discuss only the general stored energy $\WSTO$.

\section{Existing methods}
\label{SecSto}
\begin{boxexampl}[label={B:Electrostatics}]{{Electrostatic energy expressed in fields, circuits, and charges}}

Electrostatic energy $W_{\mrm{e}}$ is thoroughly treated in many classical textbooks~\cite{Jackson1999,Landau+Lifshitz1984,Feynman1965} with a clear consensus on its definition, see~\cite{Feynman1965} for a discussion. The energy $W_{\mrm{e}}$ can be expressed in three equivalent ways as
\begin{equation}\label{eq:Westatic}
	W_{\mrm{e}}=
	\frac{1}{2}\int_{\R^3}\EPS_0|\Ev(\rv)|^2\diffV 
  =\frac{1}{2}\int_{\varOmega} \phi(\rv)\rho(\rv)\diffV
	=\frac{1}{2\EPS_0}\int_\varOmega\!\int_\varOmega\frac{\rho(\rv_1) \rho(\rv_2)}{4\pi|\rv_1-\rv_2|}\diffV_1\diffV_2,
\end{equation}
where $\Ev$ denotes electric field intensity, $\phi$ electric potential and $\rho$ charge density supported in $\varOmega\subset\R^3$, see Figure~\ref{fig:FigBoundIntro}. Below, we consider a \ac{PEC} object $\varOmega$ with the total charge $\int\rho\diffV=q_{\mrm{tot}}$. From left to right, the terms in~\eqref{eq:Westatic} represent energy expressed in:
\begin{itemize}
  \item fields, where the electric energy density $\EPS_0|\Ev|^2/2$ is integrated over all space, 
  \item circuits, where a constant potential $\phi=V$ on the \ac{PEC} object is used to rewrite the energy \mbox{$W_{\mrm{e}}=Vq_{\mrm{tot}}/2=CV^2/2$} in terms of capacitance $C$,
  \item charges, where a double integral over the source region is used.
\end{itemize}
These representations offer alternative expressions and ways to evaluate the energy. Similar interpretations are observed for the electromagnetic energy discussed in Section~\ref{SecSto}. 
\end{boxexampl}

So far, we have discussed stored energy only in terms of the abstract definition in \eqref{eq:StEnDefEq3} involving the total and unobservable energies.  For practical purposes, more specific expressions are required to evaluate a system's stored energy. This Section compares many methods developed to calculate the stored energy in electromagnetic systems.  These methods vary in approach and generality, though most were motivated by the desire to calculate the Q-factor of radiating systems, as defined in \eqref{eq:qdef}.

The many attempts at defining and calculating stored energy in radiating systems can be classified and grouped in several ways, \cf the electrostatic case in Box~\ref{B:Electrostatics}.  In this section, we briefly discuss these methods using the physical quantities required in each technique as a primary distinguishing feature. All discussed methods are listed in Table~\ref{tab:methods}, where they are grouped using this convention.  Specifically, methodologies are grouped into those derived mainly from electromagnetic fields (blue color), those with energy values directly calculable from source current distributions (green color), and those which take a more abstract system-level approach (red and gray color). 

%MODIFIED TABLE
\renewcommand{\arraystretch}{1.25}
\begin{table}%
{\centering
\begin{tabular}{clcccllll}
 & \multicolumn{1}{c}{Method} & \multicolumn{3}{c}{Properties} & \multicolumn{3}{c}{Requirements} & \multicolumn{1}{c}{Reference} \\ \cmidrule{3-9}

 & & $\rv_{\mrm{ind}}$  & $W_\mathrm{sto} \geq 0$ & $\Jv$-opt   & Data & Domain & Region & \\ \toprule

\parbox[t]{3mm}{\multirow{3}{*}{\rotatebox[origin=c]{90}{Field}}} &  \cellcolor{F1}$W_{\mrm{P_\mrm{r}}}$ & \cellcolor{F1}\no & \cellcolor{F1}\ok & \cellcolor{F1}\no & \cellcolor{F1}$\Ev,\Hv$ & \cellcolor{F1}$\omega_0$ & \cellcolor{F1}$\R^3$ & \cellcolor{F1}\S\ref{sec:rdotP} \\                                           
	& \cellcolor{F2}$W_{\mrm{P}}$ & \cellcolor{F2}\ok & \cellcolor{F2}\ok & \cellcolor{F2}\no &  \cellcolor{F2}$\Ev,\Hv$ & \cellcolor{F2}$\omega_0$ & \cellcolor{F2}$\R^3$ & \cellcolor{F2}\S\ref{sec:absP}  \\                                                    
	& \cellcolor{F1}$W_{\mrm{F}}$ & \cellcolor{F1}\no & \cellcolor{F1}\no & \cellcolor{F1}\no & \cellcolor{F1}$\Ev,\Hv$ or $\Xin,\Fv$ & \cellcolor{F1}$\omega_0$ & \cellcolor{F1}$\R^3$ or Port, $S_\infty$ & \cellcolor{F1}\S\ref{sec:F2} \\ \cmidrule{2-9}
    
 \parbox[t]{3mm}{\multirow{4}{*}{\rotatebox[origin=c]{90}{Current}}} & \cellcolor{C2}$W_{\Xm'}$ & \cellcolor{C2}\ok & \cellcolor{C2}\no & \cellcolor{C2}\ok &  \cellcolor{C2}$\Zm$, $\Jm$ & \cellcolor{C2}$\omega_0$ & \cellcolor{C2}$\varOmega$ & \cellcolor{C2}\S\ref{sec:xprime} \\ 
 & \cellcolor{C1}$W_\mathrm{reac}$ & \cellcolor{C1}\ok & \cellcolor{C1}\no & \cellcolor{C1}\ok & \cellcolor{C1}$\Jv$ & \cellcolor{C1}$\omega_0$ & \cellcolor{C1}$\varOmega$ & \cellcolor{C1}\S\ref{sec:guy} \\ 
 & \cellcolor{C2}$W_{\widetilde{\Xm}'}$ & \cellcolor{C2}\ok & \cellcolor{C2}\no & \cellcolor{C2}\ok & \cellcolor{C2}$\Zm$, $\Jm$ & \cellcolor{C2}$\omega_0$ & \cellcolor{C2}$\varOmega$ & \cellcolor{C2}\S\ref{sec:statespace} \\ 
 & \cellcolor{C1}$\Wtd_\mathrm{td} \left( t_0 \right)$ & \cellcolor{C1}\ok & \cellcolor{C1}\ok & \cellcolor{C1}\no & \cellcolor{C1}$\Jtd$ & \cellcolor{C1}$t$ & \cellcolor{C1}$\varOmega$, \cellcolor{C1}$S_\infty$ & \cellcolor{C1}\S\ref{sec:td}  \\ \cmidrule{2-9}
 
\parbox[t]{3mm}{\multirow{2}{*}{\rotatebox[origin=c]{90}{System}}} & \cellcolor{S2}$W_{\mrm{Z_{in}^B}}$ & \cellcolor{S2}\ok & \cellcolor{S2}\ok & \cellcolor{S2}\no & \cellcolor{S2}$\Zin$, $\Iin$ & \cellcolor{S2}$\omega$ & \cellcolor{S2}Port & \cellcolor{S2}\S\ref{sec:circuitsynthesis} \\        
 & \cellcolor{S1}$\Wrec(t_0)$ & \cellcolor{S1}\ok & \cellcolor{S1}\ok & \cellcolor{S1} & \cellcolor{S1}$\Zin,\Iin$ & \cellcolor{S1}$\omega$ & \cellcolor{S1}Port & \cellcolor{S1}\S\ref{sec:wrec} \\ \bottomrule
           
 & \cellcolor{Q2}$\QFBW$ & \cellcolor{Q2}\ok & \cellcolor{Q2}\ok & \cellcolor{Q2}\no & \cellcolor{Q2}$\Zin$ & \cellcolor{Q2}$\omega$ & \cellcolor{Q2}Port & \cellcolor{Q2}\S\ref{sec:qfbw} \\                         
& \cellcolor{Q1}$\QZp$ & \cellcolor{Q1}\ok & \cellcolor{Q1}\ok\no & \cellcolor{Q1}\no & \cellcolor{Q1}$\Zin$ & \cellcolor{Q1}$\omega_0$ & \cellcolor{Q1}Port & \cellcolor{Q1}\S\ref{sec:qzp}  \\ \bottomrule  
       
\end{tabular}\par}      	
\caption{Methods for evaluating stored energy. Rows are grouped by the data required for its evaluation, i.e., methods derived from fields (blue), source distributions (green), and systems (red).  The final two uncolored methods are metrics not generally related to stored energy which are used for comparison purposes.  }
\label{tab:methods}
\end{table}
%%%%%%%%%%%%%%%%%%%%%%%%%%%%%%%%%%%%%%%%%%%%%%%%%%%%%%%%%%%%%

This particular division is by no means unique, and throughout this section mathematical equivalences and philosophical similarities between methods are discussed.

The data required for implementing each method are listed in the Requirements column, along with the region over which those data sets are required.  These regions are denoted using $\R^3$ to represent all space, $\varOmega$ the support of sources, $S_\infty$ the far-field sphere, and Port the port of the system.  Three salient features are indicated for each method in the Properties column.  These features are:
\begin{itemize}
\item \textit{coordinate independence}, $r_\mathrm{ind}$: A check mark in this column indicates that energy expressions are coordinate independent, \ie they are independent of an antenna's position within a coordinate system.
\item \textit{positive semi-definiteness}, $\WSTO \geq 0$: In Section~\ref{sec:SEdef} it was argued that the stored energy $\Wsto$ should always be non-negative.  A check mark in this column indicates that energies obtained by a given method obey this requirement.
\item \textit{applicability to current optimization}, $\Jv$-opt:  A check mark in this column indicates that a given formulation of stored energy can be directly applied to source current optimization, useful in determining certain physical bounds.
\end{itemize}
For the sake of simplicity, all the methods described in Section~\ref{SecSto} are presented assuming radiators made only of \ac{PEC} or assuming electric currents placed in a vacuum environment. 
All presented methods however allow generalization to non-dispersive inhomogeneous media of finite extent, although validations of such generalizations are scarce. Specific information regarding this procedure for each method is left to corresponding subsections. Similarly, certain methods may be applicable to systems containing dispersive media, though the accuracy and interpretation of results in these cases is still an open area of study.
 
\subsection{Stored energy expressed in terms of electromagnetic fields}
\begin{wrapfigure}{r}{0.45\textwidth}
%\begin{figure}
\centering
    \includegraphics{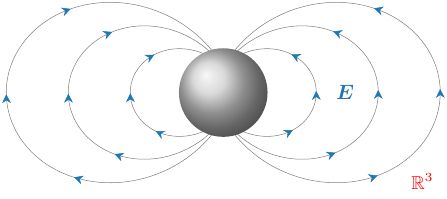} % [width=0.5\textwidth]
  \caption{Sketch of electric field intensity $\boldsymbol{E}$ generated by dominant TM$_{10}$ spherical mode.}
  \label{FigFieldConcept}
  %\end{figure}
\end{wrapfigure}
Methods derived from the fields $\Ev$ and $\Hv$ attempt to calculate stored energy \eqref{eq:StEnDefEq3} by subtracting  unobservable energy from the total energy locally at the level of electromagnetic fields around the radiator, see Figure~\ref{FigFieldConcept}. These procedures commonly allow for the definition of a local stored energy density by identifying energy in radiating fields as unobservable energy. An advantage of these methods is that they require only field quantities, not the physical structure of the radiator. However, these methods are typically computationally demanding, rendering even simple optimization tasks prohibitively expensive. 
Other common issues are the unknown form of unobservable energy within the smallest sphere circumscribing a source region $\varOmega$ (which can lead to over-subtraction \cite{Gustafsson+etal2012a}) and omission of other forms of unobservable energy such as non-radiating currents~\cite{vanBladel2007}, see also Boxes~\ref{B:UnObsWa}~and~\ref{B:UnObsWb}. In all known cases, general dispersive materials cannot be treated with these methods. The inclusion of non-dispersive materials can be made \cite{Collin+Rothschild1964,Collin1998,Yaghjian+Best2005} in all methods described in this subsection by changing $\EPS_0 \to \EPS$ and $\mu_0 \to \mu$ in the first two terms in \eqref{eq:WPr}, \eqref{eq:WP} and \eqref{eq:WF}.

The published results are dominated by analytic evaluation of the stored energy for spherical modes in the exterior region of a sphere circumscribing the radiator~\cite{Collin+Rothschild1964,Collin1998,Fante1969}. The radiated power~\eqref{eq:StEnDefEq5} expressed in the power flux and the far field are identical for this case and the classical expressions can be extended to arbitrary shapes in several ways. Here, we consider radiated energy expressed as the: power flux in the radial direction, magnitude of the power flux, and far-field amplitude, see first three rows in Table~\ref{tab:methods}.

\subsubsection{Subtraction of the radial power flow $\rvh\cdot\Pv$}
\label{sec:rdotP}
Collin and Rothschild \cite{Collin+Rothschild1964} suggested identification of radiated energy with the power flux in the radial direction to define the stored energy as
\begin{equation}
	W_{\mrm{P_\mrm{r}}} = \frac{1}{4}\int_{\R^3} \left( \EPS_0|\Ev|^2+\MUE_0|\Hv|^2-4\sqrt{\EPS_0\MUE_0}~\rvh\cdot\Pv \right) \D{V}.
\label{eq:WPr}
\end{equation}
They used this expression to evaluate the stored energy in the exterior of a sphere using mode expansions and produced explicit results on the Chu~\cite{Chu1948} lower bound, see also~\cite{Collin1998} for a time-domain extension. The expression~\eqref{eq:WPr} is non-negative and does not subtract energy for standing waves, \eg in the interior of a sphere for spherical mode expansions~\cite{Collin+Rothschild1964,Fante1969}. The main drawbacks of~\eqref{eq:WPr} are the coordinate dependence and the need for numerical integration for general fields, see~\cite{Sten+etal2001,Sten2004} for spheroidal geometries and \cite{Collardey+etal2005} for an FDTD approach.

\subsubsection{Subtraction of the magnitude of the power flow $|\Pv|$}
\label{sec:absP}
The problem with coordinate dependence in~\eqref{eq:WPr} can be resolved by subtraction of the magnitude of the power flow $|\Pv|$, \ie
\begin{equation}
	W_{\mrm{P}} = \frac{1}{4}\int_{\R^3} \left( \EPS_0|\Ev|^2+\MUE_0|\Hv|^2-4\sqrt{\EPS_0\MUE_0}|\Pv| \right) \D{V}.
\label{eq:WP}
\end{equation}
This expression for the stored energy was originally proposed in an equivalent form by Counter~\cite{Counter1948}. The expression is identical to~\eqref{eq:WPr} for fields expressed as a single spherical mode~\cite{Counter1948}. It is non-negative and less than or equal to~\eqref{eq:WPr} for general fields with a power flow in non-radial directions. The main drawback with~\eqref{eq:WP} is the numerical evaluation of the energy density over $\R^3$.  

\subsubsection{Subtraction of the far-field amplitude $|\Fv|^2$}
\label{sec:F2}
The energy of the radial component of the power flow, subtracted in the previous method~\eqref{eq:WPr}, can be expressed in the far-field amplitude $|\Fv|^2$ outside a circumscribing sphere. This leads to the formulation~\cite{Rhodes1976,Levis_AReactanceTheoremForAntennas,Yaghjian+Best2005,Fante1969,Gustafsson+Jonsson2015b,Geyi2003,Geyi2011}
\begin{equation}
	W_{\mrm{F}} = \frac{1}{4}\int\limits_{\R^3} \left( \EPS_0|\Ev|^2+\MUE_0|\Hv|^2-2\EPS_0\frac{|\Fv|^2}{|\rv|^2} \right) \D{V} = \frac{1}{4}\frac{\partial\Xin}{\partial\omega}|I_0|^2 - \frac{\Im}{2Z_0}\int\limits_{S^2}\frac{\partial\Fv}{\partial\omega}\cdot\Fv^{\ast}\D{S}
\label{eq:WF}
\end{equation}
for the stored energy, where $S^2$ denotes the unit sphere and the frequency derivatives are evaluated for a frequency independent input current $I_0$. Here, all radiated energy is subtracted and the expression makes no difference between standing and radiating waves, \eg in the interior of the smallest circumscribing sphere. Hence, the energy $W_{\mrm{F}}$ differs from $W_{\mrm{P_r}}$ by $ka P_\mrm{rad}$ for spherical modes and implies a difference of the Chu bound by $ka$, \ie $Q_{\mrm{Chu}}-ka$. Variations of~\eqref{eq:WF} exist in the literature and, \eg Rhodes \cite{Rhodes1976} suggested to use subtraction \eqref{eq:WF} only in the exterior region, keeping the total electromagnetic energy in the interior region.
A shielded power supply is also often excluded from the integration in~\eqref{eq:WF},~\cite{Yaghjian+Best2005}. This is equivalent to setting the $\Ev$ and $\Hv$ to zero in the region of the power supply.

The stored energy $W_{\mrm{F}}$ in~\eqref{eq:WF} can be rewritten using the frequency-differentiated input reactance $\Xin'$ and far field $\Fv'$ for antennas with a fixed feeding current $I_0$ using a reactance theorem~\cite{Fante1969,Rhodes1976,Yaghjian+Best2005}. This form of the stored energy is shown in the far right of \eqref{eq:WF} and simplifies the numerical evaluation from a volume integral to a surface integral. Moreover, it shows that the energy $W_{\mrm{F}}$ is coordinate dependent for non-symmetric radiation patterns~\cite{Yaghjian+Best2005,Gustafsson+Jonsson2015b}. The reactance theorem is extended to complex media in~\cite{Yaghjian+Best2005,Yaghjian2007}. The formula~\eqref{eq:WF} is also rewritten in the current density in~\cite{Gustafsson+Jonsson2015b}, see Section~\ref{sec:GV}. 

\subsection{Stored energy expressed in currents}\label{sec:current}
\begin{wrapfigure}{r}{0.3\textwidth}
\centering
    \includegraphics[width=0.80\linewidth]{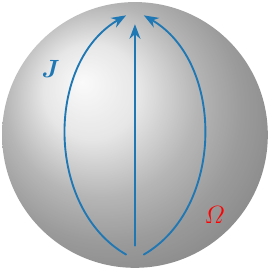} % [width=0.3\textwidth]
  \caption{Illustration of surface current of dominant TM$_{10}$ mode on a spherical shell $\varOmega$.}
  \label{FigCurrentConcept}  
\end{wrapfigure}
Several methods exist for calculating the energy stored by a source current distribution $\Jv$ placed in vacuum, see Figure~\ref{FigCurrentConcept}. These methods can be used to evaluate stored energy from any system (including materials, feeds, and ports) which can be represented by an equivalent current distribution $\Jv$. A powerful feature of this approach is an immense reduction of information needed to evaluate stored energy. Commonly, only current densities on finite surfaces are needed. These methods are also well suited for various tasks in antenna design \cite{Gustafsson+etal2016a}, since the feeding which leads to the current density $\Jv$ need not to be known. This makes it possible to determine fundamental performance bounds on antennas with given support \cite{Gustafsson+Nordebo2013,Gustafsson+etal2016a,Jelinek+Capek2017,Capek+etal2016b} or to utilize modal decomposition methods~\cite{Capek+etal2012}.

Similarly to field approaches, the methods discussed in this subsection identify radiation energy as unobservable energy. Their use for evaluation of \eqref{eq:StEnDefEq3} for lumped circuits will thus always count the entire electromagnetic energy $W_\mathrm{EM}$ regardless of the complexity of the circuit. The formulation of the methods for general dispersive materials is not well studied except for the state-space \ac{MoM} approach in Section~\ref{sec:statespace}. In the case of non-dispersive materials, electric polarization can be included in the current density $\Jv$. 

\subsubsection{Differentiated MoM reactance matrix $\Xm'$}
\label{sec:xprime}
Harrington and Mautz~\cite{Harrington+Mautz1972} proposed to use frequency differentiation of the \ac{MoM} reactance matrix 
\begin{equation}
  \WXp 
  = \frac{1}{4}\Jm^{\herm}\partder{\Xm}{\omega}\Jm
  = \frac{1}{4}\Jm^{\herm}\Xm'\Jm 
\label{eq:WXp}
\end{equation}
to estimate the stored energy. The reactance matrix is determined from the impedance matrix $\Zm=\Rm+\ju\Xm$ derived from the \ac{MoM} approximation of the \ac{EFIE}~\cite{Chew+etal2008}. The expression~\eqref{eq:WXp} is not derived in~\cite{Harrington+Mautz1972}, but is merely motivated by the analogous expression of Foster's reactance theorem for lossless systems~\cite{Harrington1968}, see also~\eqref{eq:WXinp}. The stored energy for lumped circuit networks can be determined with the formula~\eqref{eq:WXp} by substituting the \ac{MoM} impedance matrix with the lumped circuit impedance matrix, see \eqref{eq:ImpMat} and \cite{Smith1969}.

For currents in free space, the expression~\eqref{eq:WXp} is identical to the \ac{MoM} state-space approach in Section~\ref{sec:statespace} and the \ac{MoM} approximation of the stored energy expressions by Vandenbosch~\cite{Vandenbosch2010}. Hence, it also suffers from the matrix $\Xm'$ being indefinite for large structures and potentially producing negative values for the stored energy~\cite{Gustafsson+etal2012a}. The expression~\eqref{eq:WXp} is easily applied to temporally dispersive materials but is inaccurate for many cases~\cite{Gustaffson_QdisperssiveMedia_arXiv}, \cf the state-space MoM approach in Section~\ref{sec:statespace}.

\subsubsection{Reactive energy}\label{sec:GV}\label{sec:guy}
The expressions in the frequency domain introduced by Vandenbosch~\cite{Vandenbosch2010} start from the same classical idea as described by Collin and Rothschild~\cite{Collin+Rothschild1964}: the subtraction of the radiated energy density from the total energy density. However, the subtracted term is defined in a slightly different way on the basis of an energy balance equation involving the derivatives of Maxwell's laws. The resulting difference is analytically integrated over all space, yielding closed-form expressions for the reactive energy (both the electric and magnetic part) in terms of the currents flowing on the radiator. The new definition thus eliminates the coordinate dependency, resulting in the expression
\begin{multline}\label{eq:WGV}
\Wreac
=\frac{Z_0}{4\omega k}\int\limits_\varOmega\int\limits_\varOmega\bigg(  \left(k^2\Jv_1\cdot\Jv_2^{\ast} +\nabla_1\cdot\Jv_1\nabla_2\cdot\Jv^{\ast}_2\right)
\frac{\cos(kr_{12})}{4\pi r_{12}}\\
-k\big(k^2\Jv_1\cdot\Jv^{\ast}_2
-\nabla_1\cdot\Jv_1\nabla_2\cdot\Jv^{\ast}_2\big)
\frac{\sin(k r_{12})}{4\pi} \bigg) \D{V_1} \D{V_2}.
\end{multline}
This expression was later found to conform \cite{Gustafsson+Jonsson2015b} to the coordinate independent part of energy $W_{\mrm{F}}$ given by~\eqref{eq:WF}. The same expression is found also from a line of reasoning starting in time domain~\cite{Vandenbosch_RadiatorsInTimeDom1},~\cite{Vandenbosch_RadiatorsInTimeDom2}. The expression is positive semi-definite for circuits and small radiators but indefinite for larger structures~\cite{Gustafsson+etal2012a}. This method essentially can be seen as a ``transformation'' of the original field based definition~\eqref{eq:WF}, acting on all space, into a current based interpretation, acting only within the volume of the radiator. The MoM approximation of~\eqref{eq:WGV} is identical to~\eqref{eq:WXp} for the free-space case and hence~\eqref{eq:WGV} offers  a rigorous motivation for~\eqref{eq:WXp}. The first term in~\eqref{eq:WGV} is also similar to the time-domain formulation using the product of sources and potentials proposed by Carpenter in~\cite{Carpenter1989}. Moreover, Geyi presented an approximation of~\eqref{eq:WGV} for small antennas in~\cite{Geyi2003b}. This small regime formulation was also addressed in~\cite{Vandenbosch2011},~\cite{Vandenbosch_ExplicitRelationBetweenVolumeAndLowerBoundForQ}. The formulation based on~\eqref{eq:WGV} is generalized to electric and magnetic current densities in~\cite{Jonsson+Gustafsson2015,Kim2016}.

\subsubsection{State-space MoM model $\widetilde{\Xm}'$}
\label{sec:statespace}
The state-space method is based on the classical approach to define stored energy in a dynamic system, see~\eqref{eq:SysEnDef}. The stored energy for a radiating system is more complex as the dynamics are not described by the simple system in~\eqref{eq:SysEnDef}. In~\cite{TEAT-7245}, a state-space model 
\begin{equation}\label{eq:StateSpaceMoM}
  \widetilde{\Zm}\tilde{\Jm}=
  \begin{pmatrix}
    \ju\omega\MUE\Lm & \Id\\
    -\Id & \ju\omega\EPS\Cm
  \end{pmatrix}
 \begin{pmatrix}
    \Jm \\ \Um
  \end{pmatrix}	
  =\begin{pmatrix}
    \Bm \\ \Om
  \end{pmatrix}
    \Vin
\end{equation}
is derived from the \ac{MoM} impedance matrix $\Zm=\ju\omega\MUE\Lm+\Cmi/(\ju\omega\EPS)$, where $\Um$ is the voltage state and $\Vm=\Bm\Vin=\Zm\Jm$ is the excitation.
The stored energy is constructed by differentiation of the state-space reactance matrix $\widetilde{\Xm}=\Im\{\widetilde{\Zm}\}$ with respect to the frequency, \cf\eqref{eq:Zstatespace}. The resulting stored energy is identical to the $\Xm'$-formulation in Section~\ref{sec:xprime} for \ac{PEC} structures in free space and suffers from the same problem of being indefinite for larger structures. The advantage of the state-space approach is that the quadratic forms for the stored energy are derived for small structures in temporally dispersive and inhomogeneous materials.

\subsubsection{Subtraction of the radiated power in time domain} 
\label{sec:td}
The subtraction of unobservable energy~\eqref{eq:StEnDefEq3} in the form of radiation can advantageously be applied in time domain~\cite{CapekJelinekVandenbosch_EMenergiesAndRadiationQfactor}. In this paradigm the system is brought into a given state (for example time-harmonic steady state) during time $t<t_0$ and then its excitation is switched off. The system is then let to pass a subsequent transient state in which all its energy is lost via radiation and heat. With the time-dependent current density~\mbox{$\Jtd\left(t\right)$} existing in the system, which has been recorded during the entire time course, the stored energy can be calculated as
\begin{equation}
\Wtd_\mathrm{td} \left( t_0 \right) = \int\limits_{t_0}^\infty \big( \mathcal{P}_\mathrm{heat} \left( \Jtd \right)  +  \mathcal{P}_\mathrm{rad} \left( \Jtd \right) - \mathcal{P}_\mathrm{rad} \left( \Jtd_\mathrm{freeze}\right) \big) \, \mathrm{d} t,
\label{eq:WT}
\end{equation}
where~$\mathcal{P}_\mathrm{heat}$ and $\mathcal{P}_\mathrm{rad}$ are the power lost and power radiated corresponding to the lost and radiated energy~$\Wtd_\mathrm{heat}$ and~$\Wtd_\mathrm{rad}$ defined by~\eqref{eq:StEnDefEq2C},
\eqref{eq:StEnDefEq2D}, with bounding surface $S_\mathrm{far}$ located in the far field. The current density $\Jtd_\mathrm{freeze}\left(t\right)$ is defined as the current density at time $t=t_0$ artificially frozen for times $t>t_0$, \ie \mbox{$\Jtd_\mathrm{freeze}\left(t>t_0\right) = \Jtd\left(t_0\right)$}. Cycle-mean stored energy in time-harmonic case is achieved by moving time $t_0$ within one period and averaging. Note that although the power terms in \eqref{eq:WT} are evaluated for time $t>t_0$, the time retardation demands knowledge of the current density also in preceding times.

This subtraction technique closely follows the stored energy definition~\eqref{eq:StEnDefEq3} and its more detailed exposition \cite{CapekJelinekVandenbosch_EMenergiesAndRadiationQfactor} also shows that the method gives non-negative stored energy, is coordinate independent, and can subtract the radiation energy inside the smallest circumscribing sphere. Its major disadvantage is numerically expensive evaluation.
 
\subsection{Approaches using system, port, or feed}\label{sec:SysPortFeed}
\begin{wrapfigure}{r}{0.4\textwidth}
\centering
    \includegraphics{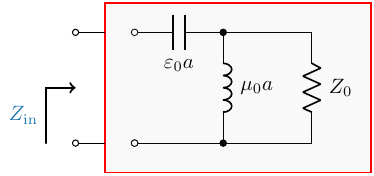}
  \caption{Synthetized circuit for dominant TM$_{10}$ mode of a spherical shell with radius $a$~\cite{Chu1948}.}
  \label{FigCircuitConcept}  
\end{wrapfigure}
System-level approaches evaluate energy storage directly from quantities available in the input/output ports of the system, see Figure~\ref{FigCircuitConcept}.  Grounded in thermodynamic principles, energy balance calculations of this kind preceded local approaches in mechanics, however, they are not commonly seen in the domain of electromagnetic stored energy evaluation. The oldest application of system-level energy quantification in electromagnetics uses circuit synthesis \cite{Chu1948,Smith1969} and is also tightly related to the concept of recoverable energy~\cite{Polevoi_MaximumExtractableEnergy}. The generality of these approaches is unprecedented as they are applicable to arbitrarily complex electromagnetic systems.  Unfortunately, this generality comes at the price of losing all physical interpretation of the unobservable energy content.  Additionally, application of these techniques require systems with well defined input ports.  This latter restriction makes these techniques inappropriate for evaluating the Q-factors of currents without a well-defined port, such as those encountered in modal decompositions and current optimization.

\subsubsection{Brune circuit synthesis}
\label{sec:circuitsynthesis}
Chu's classical antenna bound was originally derived using the  stored energy in lumped inductors and capacitors of a circuit model for the spherical modes~\cite{Chu1948}. Thal has extended this approach to hollow spheres~\cite{Thal2006} and arbitrarily shaped radiators~\cite{Thal2012}. The stored energy for arbitrarily shaped antennas can analogously be estimated from equivalent circuit networks synthesized solely from the input impedance~\cite{Gustafsson+Jonsson2015a}, where Brune synthesis~\cite{Brune1931,Wing2008} is used. Alternative synthesis methods~\cite{Wing2008} can be used but it is essential that the synthesized circuit is a reciprocal minimal representation~\cite{Willems1972b}. Non-reciprocal methods such as the minimum-phase Darlington synthesis~\cite{Smith1969,Smith1970} can be used to estimate the recoverable energy in Section~\ref{sec:wrec}. 

It is hypothesized \cite{Gustafsson+Jonsson2015a} that the Brune circuit synthesis procedure produces a circuit with minimal stored energy from all reciprocal realizations, and thus best estimates the stored energy $\Wsto$.  By definition, this means the procedure only includes the observable part of the stored energy.  Note that this is zero for the Z\"{o}bel network in Box.~\ref{B:UnObsWa}. The formulation can be used for arbitrary antennas and material models, but its application requires approximation of the input impedance $\Zin(\omega)$ as a positive-real function. This approximation is computationally difficult for electrically large antennas that require high-order rational functions.

\subsubsection{Recoverable energy}
\label{SecRec}
\label{sec:wrec} The recoverable energy ${{\Wtd}_{{\mathrm{rec}}}}\left( {{t_0}} \right)$ is defined as the maximum energy which can be extracted from a system which has been driven for times $t<t_0$ by a known set of sources \cite{1970_Day_QJMAM,Polevoi_MaximumExtractableEnergy}. In the most general sense, calculating ${{\Wtd}_{{\mathrm{rec}}}}\left( {{t_0}} \right)$ involves finding the optimal ``recovery source" \cite{Polevoi_MaximumExtractableEnergy} as a function of time $t>t_0$.  This recovery signal implicitly depends on the sources applied at times $t<t_0$ and the locations where recovery is allowed to occur.  The optimal recovery source extracts maximum energy from the system and equivalently minimizes energy lost by the system during recovery.  When both driving and recovery sources are confined to a single port as they are in many antenna systems, the task of finding the optimal recovery source is greatly simplified \cite{Direen_FundLimitsOnTerminalBehavourOfAntennas}. Given a port impedance $Z_\mathrm{c}$ and a system reflection coefficient $\varGamma(\omega)$, the recovery source (in the form of an incident voltage $u_\mathrm{in}^+(t)$) is obtained by solving
\begin{equation}
\label{Wrec01}
\mathcal{F}^{-1}\left\{ \frac{1}{Z_\mathrm{c}} \left( 1-{\left| {\varGamma \left( \omega \right)} \right|^2} \right) \right\} * u_\mathrm{in}^+(t)  = 0
\end{equation}
for times $t>t_0$, where $\ast$ denotes convolution and $\mathcal{F}^{-1}\left\{\cdot\right\}$ denotes the inverse Fourier transform.

Applying this recovery source to the antenna port, the recoverable energy is given by
\begin{equation}
{{\Wtd}_{{\mathrm{rec}}}}\left( {{t_0}} \right) = -\int\limits_{t_0}^\infty u_\mathrm{in}(t) i_\mathrm{in}(t) \diff t,
\end{equation}
where $u_\mathrm{in}$ and $i_\mathrm{in}$ are the total port voltage and current corresponding to the optimal time course $u_\mathrm{in}^+(t)$ from \eqref{Wrec01}.

For time-harmonic excitation prior to time $t_0$, the cycle-mean recoverable energy can be calculated directly in closed-form from a rational function fit of the system's input impedance~\cite{Direen_FundLimitsOnTerminalBehavourOfAntennas}. The process of approximating an antenna's input impedance as a rational function, however, suffers from the same problems as Brune synthesis for electrically large antennas. The formulation of energy ${{\Wtd}_{{\mathrm{rec}}}}$ in terms of field quantities can be found in \cite{Polevoi_MaximumExtractableEnergy} and an overview of its physical properties and more detailed exposition can be found in~\cite{2017_Schab_Arxiv_Wrec}. A first generalization of the concept to more arbitrary excitations of radiators can be found in~\cite{ZhengVandenbosch_RecoverableEnergyICEAA}.

\subsection{System-level metrics not directly derived from stored energy}\label{S:Qmetrics}
Determining the stored energy in a system is largely motivated by its approximate inverse proportionality\footnote{Often, this inverse proportionality is taken for granted. It is, however, important to stress that a strict functional relation of Q-factor based on stored energy and fractional bandwidth does not exist~\cite{Gustafsson+Nordebo2006}, and the discrepancy from the inverse proportionality can in specific cases be enormous~\cite{CapekJelinekHazdra_OnTheFunctionalRelationBetweenQfactorAndFBW}. On the other hand, in many cases, including practically all electrically small radiators, 
the inverse proportionality is almost exact.} to frequency selectivity of a single resonant system, which is most commonly described by its \ac{FBW} or Q-factor. There are however methods which attempt to evaluate Q-factor without knowledge of stored electromagnetic energy.  The most well known are the Q-factors $\QZp$ derived from the frequency derivative of an input impedance and $\QFBW$ derived directly from the fractional bandwidth of the system. Both of these methods belong to the system-based class of approaches and share those properties. For comparison purposes, both methods will be calculated alongside Q-factors derived from stored energy.

\subsubsection{Fractional bandwidth}\label{sec:qfbw}
The Q-factor $\QFBW$ is calculated directly from the fractional bandwidth $B$ as~\cite{Yaghjian+Best2005}
\begin{equation}\label{qfbw:eq01}
	\QFBW = \frac{2\refl_0}{\sqrt{1-\refl_0^2}} \frac{1}{\band_{\refl_0}},		
\end{equation}
where $\refl_0$ denotes the level of the reflection coefficient $|\refl|$ at which the \acf{FBW} $\band_{\refl_0}$
is evaluated. The relation assumes that the system is matched and tuned to resonance at the evaluation frequency, \ie $\refl(\omega) = 0$.
The most important merit of the Q-factor $\QFBW$ is its exact proportionality to fractional bandwidth. The major drawback of this method is its inability to evaluate Q-factor from data at a single frequency and its dependence on the choice of parameter $\refl_0$.

\subsubsection{Differentiated input impedance}
\label{sec:qzp}
The Q-factor $\QZp$ has been derived~\cite{Yaghjian+Best2005} from  $\QFBW$ in the limit where $\refl_0 \to 0$ and it represents the differential fractional bandwidth of the system. Similarly to $\QFBW$, it assumes the system is matched and tuned to resonance. It is most commonly defined as~\cite{Yaghjian+Best2005}
\begin{equation}
\label{qzp:eq01}
\QZp = \frac{\omega }{{2 \Rin }}\left| \frac{{\partial  \Zin }}{{\partial \omega }} \right| = \omega \left| \frac{{\partial \varGamma }}{{\partial \omega }}\right|.
\end{equation}
Alternatively, $\QZp$ can be viewed as the classical Q-factor \eqref{eq:qdef} derived from a local approximation of an input impedance by a single resonance (RLC) circuit~\cite{Yaghjian+Best2005,Ohira2016} for which relation $\QZp=Q\approx \QFBW$ holds.
The advantage of $\QZp$ over $\QFBW$ is its much simpler evaluation and its independence of the parameter $\refl_0$. However, the cost of this simplification is the loss of a direct relation to fractional bandwidth~\cite{Yaghjian+Best2005}, the possibility of predicting $\QZp = 0$~\cite{Gustafsson+Nordebo2006, CapekJelinekHazdra_OnTheFunctionalRelationBetweenQfactorAndFBW}, and the problematic interpretation in cases of closely spaced resonances~\cite{StuartBestYaghjian_LimitationsInRelatingQualityFactorToBWinAdoubleResonanceSmallAntenna}. The Q-factor $\QZp$ can also be written solely in terms of source current density \cite{CapekJelinekHazdraEichler_MeasurableQ,Gustaffson_QdisperssiveMedia_arXiv} which relates it to the Q-factor based on energies $W_{\mrm{F}}$ and $W_{\mrm{reac}}$, see Section~\ref{Sec:Compar}.

\subsection{Other methods}
The list of methods discussed above is not complete and we have intentionally selected those which follow the definition \eqref{eq:StEnDefEq3} and at the same time exhibit generality. In this subsection we briefly comment on those not explicitly treated.

First concept is that of employing angular field decomposition, identifying stored energy with the energy of the evanescent (invisible) part of the spectra \cite{Rhodes_OnTheStoredEnergyOfPlanarApertures,CollinRothschild_ReactiveEnergyInApertureFields}. A similar concept was proposed in~\cite{Thiele+etal2003} to evaluate Q-factors of electrically small dipole radiators and in~\cite{Kwon+Pozar2014} to evaluate Q-factors of arrays. This spectral decomposition method is an interesting scheme which gives important insight into the subtraction of the radiation part of unobservable energy. Its most important drawback is its applicability solely to planar radiators. A generalization to general radiators has been proposed in \cite{MikkiAntar_ATheoryOfAntennaElectromagneticearField1,MikkiAntar_ATheoryOfAntennaElectromagneticearField2}, but has not been tested.

The second concept, proposed by Kaiser \cite{Kaiser_ElectromagneticInertiaReactiveEnergyAndEnergy}, bears similarity to the time domain version of the method of Collin and Rothschild \cite{Collin1998} and claims to be its relativistic generalization. The major difference from \eqref{eq:WP} is the use of squared instead of linear subtraction which was introduced as an analogy to relativistic energy-momentum relation \cite{Bateman_TheMathematicalAnalysisOfElectricalAndOpticalWaveMotion,Kaiser_ElectromagneticInertiaReactiveEnergyAndEnergy}. The merit of this concept is positive semi-definiteness, coordinate independence, and the capability to deliver a local stored energy density. In canonical cases it leads to stored energy values \cite{CapekJelinek_VariousInterpretationOfTheStoredAndTheRadiatedEnergyDensity} very close to \eqref{eq:WP}, but its testing in more general scenarios is not available. 

The last presented concept is based on a fact that the stored energy in a lossless network can be determined by differentiation of the input reactance $\Xm_{\mrm{in}}$ or susceptance $\mat{B}_{\mrm{in}}$~\cite{Harrington1968} as
\begin{equation}
  W_{\Xm_{\mrm{in}}'} 
  = \frac{1}{4}\Jm_{\mrm{in}}^{\herm}\partder{\Xm_{\mrm{in}}}{\omega}\Jm_{\mrm{in}}
  %= \frac{1}{4}\Jm_{\mrm{in}}^{\herm}\Xm_{\mrm{in}}'\Jm_{\mrm{in}} 
  \quad\text{and}\quad
  W_{\mat{B}_{\mrm{in}}'} 
  = \frac{1}{4}\Vm_{\mrm{in}}^{\herm}\partder{\mat{B}_{\mrm{in}}}{\omega}\Vm_{\mrm{in}},
  %= \frac{1}{4}\Vm_{\mrm{in}}^{\herm}\Xm_{\mrm{in}}'\Vm_{\mrm{in}} 
\label{eq:WXinp}
\end{equation}
respectively.
This formula is related to the Foster's reactance theorem~\cite{Foster1924} where a positive energy implies a positive slope of the reactance. The input resistance of antennas is, however, non-zero and the approximation~\eqref{eq:WXinp} is hence generally inadequate. This is also concluded from~\eqref{eq:WF},  as~\eqref{eq:WXinp} neglects the far-field term in~\eqref{eq:WF}. Moreover, it is necessary to include the input resistance to accurately estimate the fractional bandwidth as shown by $\QZp$ expression in~\eqref{qzp:eq01}. Although the expression~\eqref{eq:WXinp} has the same form as the differentiated reactance matrices in Sections~\ref{sec:xprime} and~\ref{sec:statespace} there are substantial differences. It is sufficient to know only the input-output relation for the lossless system in~\eqref{eq:WXinp} whereas~\eqref{eq:WXp} requires knowledge of the internal dynamics of the system. 
\section{Analytic and numerical comparisons}\label{sec:ANCompar}
In this section, two classes of comparisons are made between the methods described in the preceding section.  First, we study the analytic relation between some methods under certain specific conditions.  Following that, numerical examples are presented where the Q-factor of driven antennas are calculated and compared.

\subsection{Analytical comparison of various methods}
\label{Sec:Compar}
When methods from Table~\ref{tab:methods} are applied to fields and currents generated by \ac{PEC} structures operating in the quasi-static limit where radiation is negligible, the stored energy predicted by them reduces to the electro- and magnetostatic expressions, see Box~\ref{B:Electrostatics}. They however start to differ for electrically larger structures. Here, the methods are analytically compared for canonical cases such as spherical geometries, \ac{PEC} structures, and single-resonance models.

Spherical modes have dominated evaluation of stored energy and Q-factors since the publication by Chu~\cite{Chu1948}. Collin and Rothschild~\cite{Collin+Rothschild1964}, see Section~\ref{sec:rdotP}, presented closed form expressions of the Q-factor and stored energy $W_{\mrm{P_r}}$ for a single radiating spherical mode outside a sphere with radius $a$. Comparing the definitions of the methods in Table~\ref{tab:methods} for this case reveals the identities
\begin{equation}
  W_{\mrm{P_r}} = W_{\mrm{F}}+\frac{a}{c_0} P_{\mrm{rad}} = W_{\mrm{P}} = W_{\mrm{Z_{in}^B}},
\label{eq:CompSphChu}
\end{equation}
where the difference with $a P_{\mrm{rad}}/c_0$ ($ka$ for the Q-factor) for the subtracted far-field expression $W_{\mrm{F}}$ originates from the subtraction of the radiated power inside of the sphere in~\eqref{eq:WF} and the equality for the Brune circuit follows from the circuit model of the spherical modes~\cite{Chu1948}. Thal~\cite{Thal2006} analyzed
the corresponding case with electric currents by inclusion of the stored energy in standing waves inside the sphere. This case is identical to~\eqref{eq:CompSphChu} for the field-based methods but with an added connection to $W_{\mrm{reac}}$, \ie  
\begin{equation}
  W_{\mrm{P_r}} = W_{\mrm{F}}+\frac{a}{c_0} P_{\mrm{rad}} = W_{\mrm{P}}
  =W_{\mrm{reac}}+\frac{a}{c_0} P_{\mrm{rad}},
\label{eq:CompSphThal}
\end{equation}
where the spherical mode expansion in~\cite{Gustafsson+Jonsson2015b} is used for $W_{\mrm{reac}}$ in~\eqref{eq:WGV}. The identity~\eqref{eq:CompSphThal} can be generalized to arbitrary electric current densities on the sphere with exception for $W_{\mrm{P}}$.

When stored energy $W_{\mrm{F}}$ given by \eqref{eq:WF} is written as a bilinear form of source current density \cite{Gustafsson+Jonsson2015b}, it relates to energy \eqref{eq:WGV} as \mbox{$W_{\mrm{F}} = W_{\mrm{reac}} + W_{\mrm{coord}}$}, where coordinate-dependent term $W_{\mrm{coord}}$ is given by \cite[Eq. 26]{Gustafsson+Jonsson2015b}. The coordinate dependent part vanishes in the important case of equiphase current densities, \ie $|\Jm^{\trans}\Jm|=\Jm^{\herm}\Jm$,  which appear as a result of characteristic mode decomposition~\cite{Capek+etal2012}, minimum Q-factor modes~\cite{Capek+etal2016b}, and often approximately for small self-resonant antennas. The equiphase case is also related to differentiation of the input admittance~\eqref{eq:WXinp} for a fixed voltage source~\cite{Gustaffson_QdisperssiveMedia_arXiv} revealing the following connection between the field, current, and port based methods:
\begin{equation}
  W_{\mrm{F}} = W_{\mrm{reac}} = |W_{\mat{B}_\mrm{in}'}| \approx Q_{\mrm{Z}'}\frac{P_{\mrm{rad}}}{\omega},
\label{eq:CompEqPhase}
\end{equation}
where the final step is valid for self-resonant cases for which the change of reactance dominates over the resistance. 

The \ac{MoM} discretized version of~\eqref{eq:WGV} for \ac{PEC} structures is also identical to the differentiated reactance matrix~\eqref{eq:WXp} and the state-space MoM~\eqref{eq:StateSpaceMoM}, \ie
\begin{equation}
  W_{\mrm{reac}}=W_{\Xm'}=W_{\widetilde{\Xm}'}.
\label{eq:CompPEC}
\end{equation}
This equality is used for the presented numerical results in Section~\ref{sec:NumComp}, where the energy $W_{\mrm{reac}}$ is used to indicate all three methods in~\eqref{eq:CompPEC}. 

Finally, the system methods agree for single-resonance RLC circuit networks
\begin{equation}
  Q_{\mrm{Z_{in}^B}}=Q_{\mrm{rec}}=Q_{\mrm{Z}'}\approx Q_{\mrm{FBW}},
\label{eq:Qcomparison}
\end{equation}
where the subscripts used are the same as for corresponding energies.

\iftoggle{fullpaper}{%
\begin{figure}%
{\centering
 \begin{subfigure}{0.45\textwidth}
        \includegraphics[width=\textwidth]{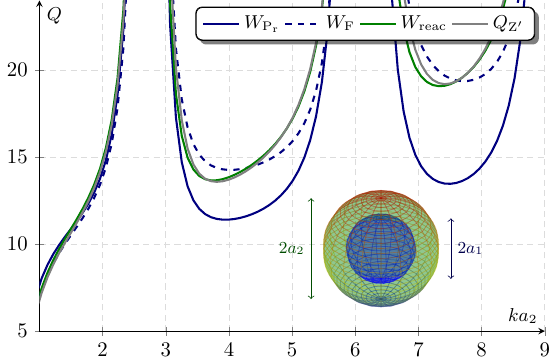}
          \caption{}
          \label{fig:ConSpha}
      \end{subfigure}
      \hspace{5mm}
      \begin{subfigure}{0.45\textwidth}
        \includegraphics[width=\textwidth]{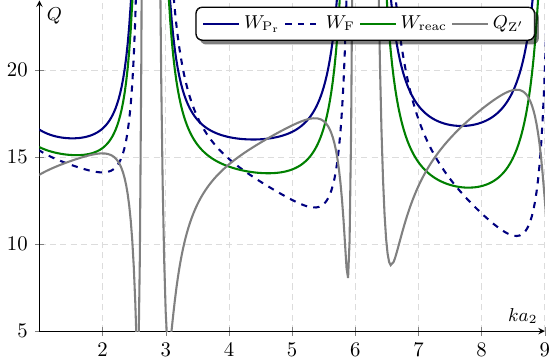}
          \caption{}
          \label{fig:ConSphb}
      \end{subfigure}
      \par}
\caption{Q-factors for concentric spherical current shells radiating the spherical TM$_{01}$ mode: a) $ka_1=0.75$ and $J_2=(0.5+0.1\ju)J_0$, b) $ka_1=0.3$ and $J_2=\ju J_0$.}%
\label{fig:ConSph}%
\end{figure}}{}
The above comparison suggests that the proposed methods agree for many cases. However, the identities are based on specific assumptions and discarding the imposed restrictions on the geometries, equiphase currents, and single resonance can produce very different estimates of the stored energy. 
\iftoggle{fullpaper}{%
As an example, we generalize the single mode case~\eqref{eq:CompSphThal} to single electric dipole mode (TM$_{01}$) originating from electric currents at two spherical shells with radii $a_1$ and $a_2>a_1$. Let the inner current have amplitude $J_1$ and normalize the outer current amplitude with $J_0$ such that $J_2=J_0$ cancels the radiation from the inner surface. This non-radiation current has no dissipated power and hence an infinite Q-factor. Lowering the amplitude to $J_2=0.5 J_0$ increases total the radiation as only half of the radiated field is canceled. Figure~\ref{fig:ConSpha} depicts the case $ka_1=0.75$ with $J_2=(0.5+0.1\ju) J_0$, where the small imaginary part is added to invalidate the equiphase identity~\eqref{eq:CompEqPhase}. In the figure, we observe that $Q_{\mrm{F}}\approx Q_{\mrm{reac}}\approx Q_{\mrm{Z}'}$ as expected from~\eqref{eq:CompEqPhase} as the current is approximately equiphase. The Q-factors from the subtracted power flow~\eqref{eq:WPr} and~\eqref{eq:WP} are substantially lower than the other Q-factors around $ka_2\approx 4$ and $ka_2\approx 8$. This is contrary to the expectation from the single mode case~\eqref{eq:CompSphThal} and can be explained by the power flow between the spherical shells that is not subtracted by the far field in~\eqref{eq:WF}. The effects on the Q-factors of an increased phase shift between the current is depicted in Figure~\ref{fig:ConSphb}, where $ka_1=0.3$ and $J_2=\ju J_0$ is used. Here, all considered methods produce different results. These simple examples illustrate the challenges to define stored energy and that the challenge increases with the electrical size of the object and phase variation of the current.}{}

\subsection{Numerical comparison of various methods}\label{sec:NumComp}
Numerical results for different antenna types are presented in this section. The examples are: a center fed cylindrical dipole, an off-center fed cylindrical dipole, a strip folded dipole, and a Yagi-Uda antenna. The tuned Q-factor~\eqref{eq:tunedqdef} is chosen as an appropriate measure to compare the different methods, as it is only a renormalization of the stored energy along with an addition of a known tuning energy, see Section \ref{sec:SEdef}. This permits us to compare and contrast methods for evaluating the stored energy with the methods in Section~\ref{S:Qmetrics} which only calculate the tuned Q-factor, such as $\QZp$ and $\QFBW$. All example structures are modeled as \ac{PEC} in free space and are each fed by a single delta-gap voltage source. In this case many of the methods described in Section~\ref{SecSto} are formally equivalent, see Section \ref{Sec:Compar}. Hence, only one representative of each such group is presented here. Each method follows the notation introduced in Table~\ref{tab:methods}. The frequency axis of all plots is expressed in the dimensionless quantity $ka$, where $a$ is the radius of the smallest sphere that circumscribes each antenna. The Q-factor $\QFBW$ has been calculated at the level $\varGamma_0=1/3\approx-10\unit{dB}$ in~\eqref{qfbw:eq01}. 

\subsubsection{Center fed cylindrical dipole}\label{sec:CenCylDip}
\begin{figure}[]
\centering
\includegraphics[width=0.45\textwidth]{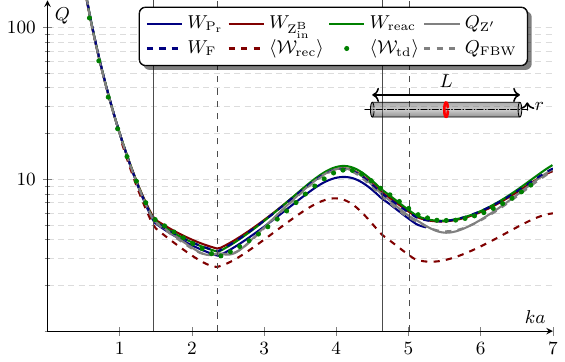}
\caption{Q-factors of a hollow cylindrical dipole of length $L$ and radius $r =L/200$, fed at its center. The gray solid and dashed vertical lines denote resonance and anti-resonances of the antenna.} 
\label{fig:Centerdipole}  
\end{figure}

Figure~\ref{fig:Centerdipole} depicts the Q-factors calculated by the methods discussed in Section~\ref{SecSto} for a hollow cylindrical dipole. All the methods agree well for low $ka$ values, which are typical dimensions for electrically small antennas. The methods start to diverge for electrically larger structures, when $ka\gg 1.5$. It should be noted that the relative difference in Q-factor is very small, even for larger structures. The only major divergence is the Q-factor from the recoverable energy $W_{\mrm{rec}}$ which predicts significantly lower values than the other methods for $ka > 3$. This, however, is to be expected as the recoverable energy is the lower bound to the stored energy, see~\eqref{eq:StEnDefEq4}.

\iftoggle{fullpaper}{%
\subsubsection{Off-center fed cylindrical dipole}\label{sec:OffCenCylDip}
The dipole examined here is identical to the center fed dipole in Section~\ref{sec:CenCylDip} except that its feeding point is shifted by a distance \mbox{$l = 0.23L$} from the center. This gives rise to a phase shift which changes the stored energy and Q-factor. If we compare Figures~\ref{fig:Centerdipole} and~\ref{fig:OffCenterDipole} we see that the Q-factors fluctuate much more than observed in the center fed dipole. However, the Q-factors retain the same behavior with respect to each other as for the center fed dipole for most of the simulated interval. They predict essentially the same results for low values of $ka$ and diverge slightly for $ka>1.5$. However, around $ka=6.2$ the Q-factor $\QZp$  has a dip which is not mimicked by the other methods. The recoverable energy $W_{\mrm{rec}}$ predicts lower values of Q-factor than the other methods but seems to follow the behavior of the curves with smaller fluctuations.
\begin{figure}[]
\centering
\includegraphics[width=0.45\textwidth]{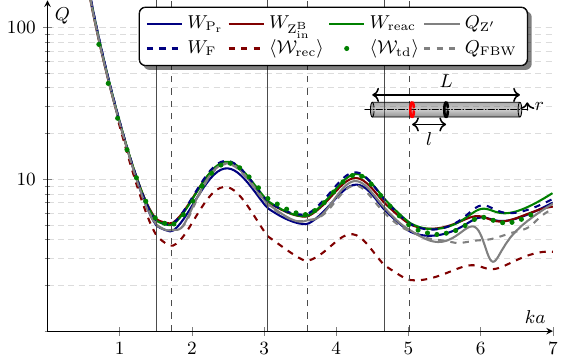}
\caption{Q-factors for a hollow cylindrical dipole of length $L$ and radius $r =L/200$, with an off-center feed $l = 0.23L$ from the center. The gray solid and dashed vertical lines denote resonance and anti-resonances of the antenna.} 
\label{fig:OffCenterDipole}  
\end{figure}
}{}

\subsubsection{Strip folded dipole}\label{sec:StripFoldDip}
\begin{figure}[]
\centering
\includegraphics[width=0.45\textwidth]{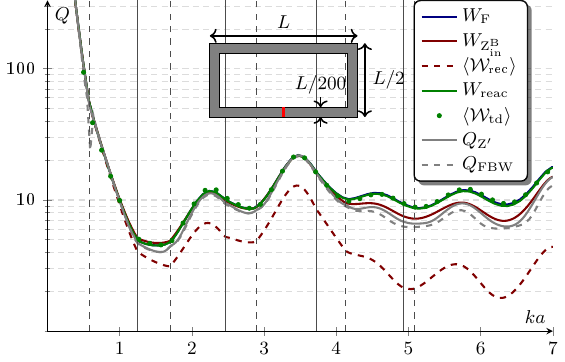}
\caption{Q-factors for a folded strip dipole of circumscribing dimensions $L \times L/2$, with strip width $L/200$. The gray solid and dashed vertical lines denote resonance and anti-resonances of the antenna.} 
\label{fig:FoldDip}  
\end{figure}

In Figure~\ref{fig:FoldDip}, Q-factors are depicted for a folded strip dipole. Due to computational complexity the subtraction of the power flow $|\Pv|$, the energy $W_{\mrm{P}}$ has not been calculated for this example.   With exception of recoverable energy, the depicted methods shown agree well for $ka<4$, above this point the Q-factors $\QZp$ and $\QFBW$ start to diverge from the other methods. 

\subsubsection{Yagi-Uda}\label{sec:YagiUda}
\begin{figure}[]
\centering
\includegraphics[width=0.45\textwidth]{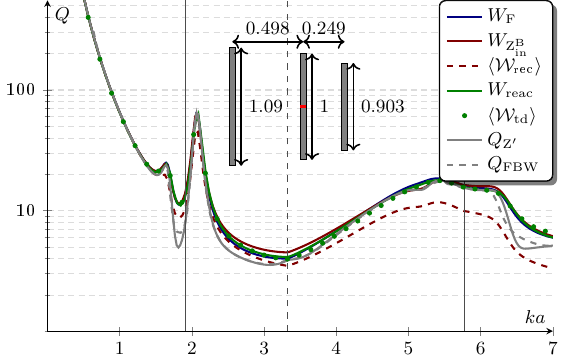}
\caption{Q-factors for a Yagi-Uda antenna specified in the upper right corner of the figure. All the dimensions of the Yagi-Uda antenna are normalized to the center dipole length $L$. The elements have been modeled as strips of width $L/200$. The gray solid and dashed vertical lines denote resonance and anti-resonances of the antenna.} 
\label{fig:YagiUda}  
\end{figure}

Figure~\ref{fig:YagiUda} depicts Q-factors calculated for a Yagi-Uda antenna, again the subtraction of the power flow, $|\Pv|$ has not been calculated due to computational complexity. All methods presented agree well over the entire interval, excluding a small dip from Q-factor $\QZp$ at $ka=1.8$ and some small divergence at $ka>6$. This can be explained by the off resonance behavior of the Yagi-Uda antenna. When the parasitic elements are no longer active, the antenna essentially behaves as a center-fed dipole. Because of this simple behavior the relative difference between the methods becomes very small.

\iftoggle{fullpaper}{%

\section{Applications}
\label{sec:Applications}
Stored energy for radiating systems was initially used by Chu~\cite{Chu1948} to derive his classical antenna bounds for spherical shapes. Bounds have continued to be a major driving force for research into stored energy as antenna designers are, naturally, interested in how good their antennas are and how far they are from the optima~\cite{Sievenpiper+etal2012,Volakis+etal2010,Gustafsson+etal2015b}. The Chu bound was originally derived with a circuit model for spherical modes~\ref{sec:circuitsynthesis}, see also~\cite{Thal2006,Thal2009,Thal2012}. The model was reformulated in fields~\eqref{eq:WPr} by Collin and Rothschild~\cite{Collin+Rothschild1964} and subsequently refined in~\cite{Fante1969,McLean1996,Yaghjian+Best2005}, see~\cite{Volakis+etal2010,Gustafsson+etal2015b,Sievenpiper+etal2012} for an overview. Formulations as optimization problems has generalized the classical bounds on the Q-factor to a multitude of problems formulated as combinations of stored energy, radiated fields, induced currents, and losses~\cite{Gustafsson+Nordebo2013,Gustafsson+etal2016a,Jelinek+Capek2017,Capek+etal2016b}. Many problems are formulated as convex optimization problems~\cite{Gustafsson+Nordebo2013,Gustafsson+etal2016a,Tayli+Gustafsson2016,Gustafsson+etal2015a,Capek+etal2016b} which are efficiently solved with standard algorithms. Here, it is essential that the quadratic forms for the stored energy are positive semidefinite, see Table~\ref{tab:methods}. Unfortunately, several presented methods are indefinite for electrically large structures. This restricts the problems to sub-wavelength structures where the expressions are positive semidefinite. Apart from convex optimization and considering mainly sub-wavelength radiators, other techniques like parameter sweeps~\cite{Vandenbosch2011,Vandenbosch_ExplicitRelationBetweenVolumeAndLowerBoundForQ}, polarizabilities~\cite{Jonsson+Gustafsson2015,Yaghjian+Stuart2010,Yaghjian+etal2013}, or modal decomposition~\cite{Kim2016,Chalas+etal2016,Capek+Jelinek2016,Jelinek+Capek2017} can be applied to determine bounds.
  
Although stored energy has so far mainly been used to determine physical bounds, stored energy has great potential to be an important concept also for an antenna design. The results by Chu~\cite{Chu1948} showed that small antennas are dipole radiators and the explicit shape of the current distribution can give insight to design. Thal~\cite{Thal2006} showed how the stored energy in the interior of a sphere contributes~\cite{Best2004,Kim+etal2010}. The importance of the polarizability and its associated charge separation was shown in~\cite{Gustafsson+etal2007a,Yaghjian+etal2013}. With the current-based formulations in Section~\ref{sec:current} and optimization of the current distribution we get suggestions for optimal currents for many antenna parameters~\cite{Gustafsson+etal2012a,Gustafsson+Nordebo2013,Gustafsson+etal2016a,Jelinek+Capek2017,Capek+etal2016b}.

Another direction from which the problem of minimization of Q-factor was attacked is characteristic mode theory \cite{HarringtonMautz_TheoryOfCharacteristicModesForConductingBodies} as it provides favorable separation of reactive stored energy~\eqref{eq:WGV}, constituting thus modal Q-factors for arbitrary bodies~\cite{Capek+etal2012}. Mixing rules similar to those used with spherical modes can be applied, leading to approximative, but straightforward rules for fundamental bounds on Q-factor of arbitrarily-shaped radiators. Stored energy expressions are also used to construct new type of modes with properties differing from those of characteristic modes. Energy modes formed from eigenvalue problems involving matrix $\Xm'$ in~\eqref{eq:WXp} were introduced in~\cite{Harrington+Mautz1972}. These types of modes are also useful to determine and interpret the physical bounds discussed above~\cite{Capek+Jelinek2016,Gustafsson+etal2016a}. Moreover, as these modes are real-valued many of the proposed expressions for stored energy agree~\eqref{eq:CompEqPhase} and the resulting Q-factor is also a good estimate of the fractional bandwidth for single mode antennas.

Stored energy can also be used to simplify some antenna optimization by replacing simulations over a bandwidth with a single frequency calculation of the Q-factor~\cite{Cismasu+Gustafsson2014a}. This single frequency optimization increases the computational efficiency but is restricted to narrow band cases. A typical representative of an application which can enjoy this approach is a design and optimization of \ac{RFID} tags with minimal mutual coupling \cite{VenaEtAl_FullyPrintableChiplessRFIDTag, PolivkaEtAl_RCSUshapedRFID}.

}{%
  % electronic
}

\section{Summary}
\label{sec:discussion}
A definition of stored energy in a general electromagnetic system was proposed and discussed using the concept of unobservable energy. Various aspects of subtracting the unobservable energy have been pointed out in the examples of Z\"{o}bel's network, matched transmission lines, and, most importantly, radiating structures. It has been shown that a majority of the well-established concepts for evaluating stored energy in radiating systems can be categorized into three different groups -- whether they used field quantities, source currents, or rely solely on knowledge in system as a whole without possibility to probe its internal structure. An important outcome of this paper is understanding that all existent concepts, in fact attempt do define unobservable energy. Nevertheless, the common association of unobservable energy purely with radiated energy is insufficient. By the proposed definition, the unobservable energy represents the difference between the total electromagnetic energy $\WEM$ and the stored energy $\WSTO$ so that it contains the energy of all unobservable states.

Careful analysis of the presented results revealed good agreement between all evaluated methods for equiphased currents and electrically small ($ka < 1.5$) antenna structures, though simple analytically-constructed examples and larger objects revealed significant disagreements. The systematic difference between recoverable energy $\WREC$ and stored energy $\WSTO$ is due to reciprocity of the resulting realizations. While the recoverable energy allows for non-reciprocal circuits, the stored energy approaches, as illustrated by Brune synthesis, deal with reciprocal systems only. Taking $\QFBW$ as reference measure of fractional bandwidth, it is obvious that the Q-factor resulting from recoverable energy considerably overestimates the fractional bandwidth. The other presented methods have much better agreement with fractional bandwidth. However, from this point of view, the best predictor of bandwidth potential is Q-factor~$\QZp$, but only when the system under study can be approximated as a single resonance system.

For practical aspects of stored energy evaluation, the method evaluating energy $\Wreac$ or, alternatively, energy~$\WXp$, gives precise approximation of stored energy for electrically small structures, offers simple implementation, and, in addition, is fully compatible with present approaches to minimization of Q-factor like convex optimization and pixeling. Whenever negative values of stored energy could be an issue, an alternative method, possibly Brune synthesis, is recommended since the breaking point at which stored energy $\Wreac$ fails is not exactly known. As confirmed by all treated examples, Brune synthesis is capable of distilling the maximum amount of unobservable energy from the total energy, thus surpassing other contemporary approaches. However, complications in performing Brune synthesis for electrically large antennas may be an obstacle limiting its application.

Though many researchers have contributed to the study of stored energy with corresponding indisputable achievements, several fundamental questions remain open. The missing proof of the minimal reciprocal realizations generated by Brune synthesis as well as closely related reformulation of this circuit synthesis in terms of the electromagnetic quantities,  may open the final stage to explicit, coherent, and exact definition and evaluation of unobservable energy.  Additionally, further work is needed on the calculation, verification, and interpretation of stored energy in general dispersive media.

\iftoggle{fullpaper}{%
  % using paper
\appendices

\section{Stored energy in dispersive media}\label{app:SEdefDisp}
The definition in Section~\ref{sec:SEdef} covers antennas in a non-dispersive background. Consider instead a radiator embedded in an isotropic dielectric material described by a Lorentz dispersion model 
\begin{equation}\label{eq:ApnLorMat}
\frac{{{\partial ^2}{\Ptd}}}{{\partial {t^2}}} + \Gamma \frac{\partial \Ptd}{{\partial t}} + \omega _{\mathrm{r}}^2{\Ptd} = {\varepsilon _0}\omega _{\mathrm{p}}^2{\Etd},
\end{equation}
where $\Ptd$ is the polarization, $\Gamma$ is the loss factor, $\omegar$ is the resonance frequency of the material, and $\omegap$ is the coupling constant \cite{Jackson1999}. If we divide the energy analogously to~\eqref{eq:StEnDefEq1}, the material properties influence the heat and total energy terms \cite{Ruppin_ElectromagneticEnergy}. The new heat term reads
\begin{equation}
\label{eq:AppHeat}
\Wtd_\mathrm{heat} \left(t_0\right) = \int\limits_{-\infty}^{t_0} \int\limits_V \left( \SIGMA \left| \Etd \right|^2 + \frac{\Gamma }{{{\varepsilon _0}\omega _{\mathrm{p}}^2}}{\left| {\frac{{\partial {\Ptd}}}{{\partial t}}} \right|^2} \right) \D{V} \D{t},
\end{equation}
and the total energy read
\begin{equation}
\label{eq:AppStEnDef}
\Wtd_\mathrm{EM} \left(t_0\right) = \frac{1}{2} \int\limits_V \left(\EPS_0 \left| \Etd \right|^2 + \MUE_0 \left| \Htd \right|^2 + \frac{1}{{{\varepsilon _0}\omega _{\mathrm{p}}^2}}\left[ {{{\left| {\frac{{\partial {\Ptd}}}{{\partial t}}} \right|}^2} + \omega _{\mathrm{r}}^2{{\left| {\Ptd} \right|}^2}} \right] \right) \D{V}.
\end{equation}
The stored energy definition \eqref{eq:StEnDefEq3} still applies, but the dispersion generally rise the energy of unobservable states. The subtraction of unobservable energy states becomes especially problematic in dispersive background since in a such case far field is no longer well defined and many classical methods break down. System based methods, see Table~\ref{tab:methods}, and engineering metrics $\QZp$ and $\QFBW$ are unaffected, in principle, but, in certain cases, they are more likely to predict unphysical results, see~\cite{Gustaffson_QdisperssiveMedia_arXiv}.

\ifCLASSOPTIONcaptionsoff
\newpage
\fi
\bibliographystyle{IEEEtran}
\bibliography{total,references_LIST_UpToDate}

\begin{IEEEbiographynophoto}{Kurt Schab} received the B.S. degree in electrical engineering and physics from Portland State University, Portland, OR, USA, in 2011, and the M.S. and Ph.D. degrees in electrical engineering from the University of Illinois at Urbana-Champaign, Champaign, IL, USA, in 2013 and 2016, respectively.  Currently, he is a postdoctoral research fellow at North Carolina State University. His research interests include electromagnetic theory, optimized antenna design, and numerical methods in electromagnetics.
\end{IEEEbiographynophoto}

\begin{IEEEbiographynophoto}{Lukas Jelinek}
received his Ph.D. degree from the Czech Technical University in Prague, Czech Republic, in 2006. In 2015 he was appointed Associate Professor at the Department of Electromagnetic Field at the same university. His research interests include wave propagation in complex media, general field theory, computational electromagnetics and optimization.
\end{IEEEbiographynophoto}

\begin{IEEEbiographynophoto}{Miloslav Capek}
(S'09, M'14) received his Ph.D. degree from the Czech Technical University in Prague, Czech Republic, in 2014. In 2017 he was appointed Associate Professor at the Department of Electromagnetic Field at the same University. He leads the development of the AToM (Antenna Toolbox for Matlab) package. His research interests are in the area of electromagnetic theory, electrically small antennas, numerical techniques, fractal geometry and optimization. He authored or co-authored over 65 journal and conference papers. He is member of Radioengineering Society, regional delegate of EurAAP, and Associate Editor of Radioengineering.
\end{IEEEbiographynophoto}

\begin{IEEEbiographynophoto}{Casimir Ehrenborg}  received his M.Sc. degree in engineering physics  from Lund University, Sweden, in 2014. He is currently a Ph.D. student in the Electromagnetic Theory Group, Department of Electrical and Information Technology, Lund University. In 2015, he  participated in and won the IEEE Antennas and  Propagation Society Student Design Contest for his body area network antenna design. His research interests include antenna theory, phase and radiation centers, as well as physical bounds.
\end{IEEEbiographynophoto}

\begin{IEEEbiographynophoto}{Doruk Tayli}
received his B.Sc. degree in Electronics Engineering from Istanbul Technical University and his M.Sc. in degree in Communications Systems from Lund University, in 2010 and 2013, respectively. He is currently a Ph.D. student at Electromagnetic Theory Group, Department of Electrical and Information Technology at Lund University. His research interests are Physical Bounds, Small Antennas and Computational Electromagnetics.
\end{IEEEbiographynophoto}

\begin{IEEEbiographynophoto}{Guy A. E. Vandenbosch}
is a Full Professor at KU Leuven, Belgium. His interests are in the area of electromagnetic theory, computational electromagnetics, planar antennas and circuits, nano-electromagnetics, EM radiation, EMC, and bio-electromagnetics.  His work has been published in ca. 265 papers in peer reviewed international journals and has led to ca. 365 presentations at international conferences. He is a former chair of the IEEE AP/MTT Benelux Chapter and currently, he leads the Working Group on Software within EuRAAP. Dr. Vandenbosch is a fellow of the IEEE since January 2013.\end{IEEEbiographynophoto}

\begin{IEEEbiographynophoto}{Mats Gustafsson}
received the M.Sc. degree in Engineering Physics 1994, the Ph.D. degree in Electromagnetic
Theory 2000, was appointed Docent 2005, and Professor of Electromagnetic Theory 2011, all from Lund University, Sweden. 

He co-founded the company Phase holographic imaging AB in 2004. His research interests are in scattering and antenna theory and inverse scattering and imaging. He has written over 80 peer reviewed journal papers and over 100 conference papers. Prof. Gustafsson received the IEEE Schelkunoff Transactions Prize Paper Award 2010 and Best Paper Awards at EuCAP 2007 and 2013. He served as an IEEE AP-S Distinguished Lecturer for 2013-15.
\end{IEEEbiographynophoto}

%\section*{Biography -- About seven brave men attacking the problem of stored energy $:)$}
\vspace{1cm}
\begin{center}
  \includegraphics[width=15cm]{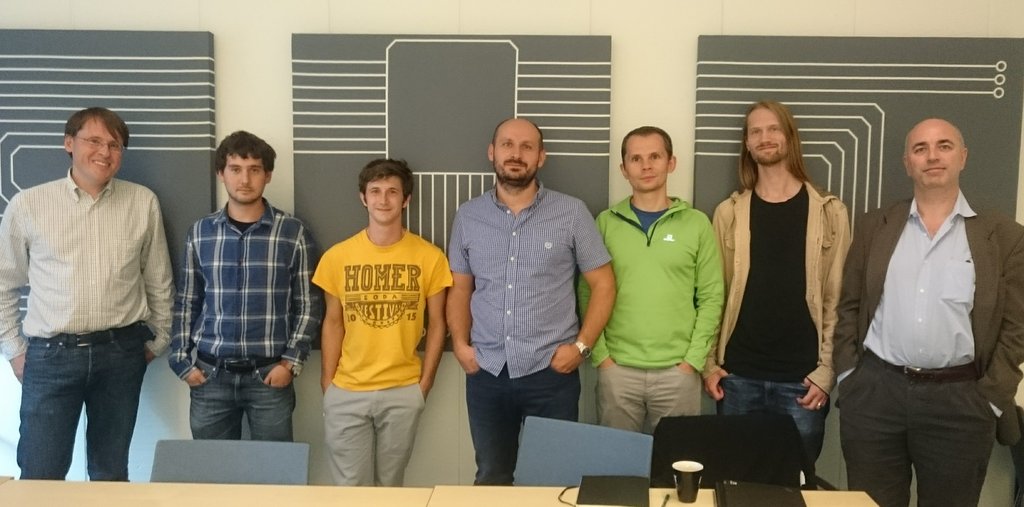}\\
  \footnotesize{L to R: Mats Gustafsson, Doruk Tayli, Kurt Schab, Lukas Jelinek, Miloslav Capek, Casimir Ehrenborg, Guy Vandenbosch}
\end{center}

}{%
  % electronic
\ifCLASSOPTIONcaptionsoff
\newpage
\fi
\bibliographystyle{IEEEtran}
\bibliography{total,references_LIST_UpToDate}
  
}

\end{document}